\documentclass[letterpaper]{article} 
\usepackage{aaai2026}  
\usepackage{times}  
\usepackage{helvet}  
\usepackage{courier}  
\usepackage[hyphens]{url}  
\usepackage{graphicx} 
\usepackage{natbib}  
\usepackage{caption} 
\usepackage{multirow}
\usepackage{booktabs}
\usepackage{colortbl}
\usepackage{longtable}
\usepackage{tabularx}
\usepackage{algorithm}
\usepackage{algorithmic}
\usepackage{svg}

\usepackage{newfloat}
\usepackage{listings}
\DeclareCaptionStyle{ruled}{labelfont=normalfont,labelsep=colon,strut=off} 
\floatstyle{ruled}
\newfloat{listing}{tb}{lst}{}
\floatname{listing}{Listing}
\usepackage{xcolor}

\title{The Perils of Agency: How Developers Perceive, Prioritize, and Address Risks in Agentic AI Products}

\author{
Hao-Ping (Hank) Lee\textsuperscript{\rm 1}\footnote{Work done while at IBM Research.},
Jessica He\textsuperscript{\rm 2},
David Piorkowski\textsuperscript{\rm 3}\footnotemark[\value{footnote}],
Thomas Serban von Davier\textsuperscript{\rm 1},
Jodi Forlizzi\textsuperscript{\rm 1},
Sauvik Das\textsuperscript{\rm 1}
}
\affiliations{
\textsuperscript{\rm 1}Carnegie Mellon University\\
\textsuperscript{\rm 2}IBM Research\\
\textsuperscript{\rm 3}Alinia\\
}

\begin{document}

\maketitle

\begin{abstract}
    Agentic AI systems act autonomously, use tools, adapt to context, and operate in complex real-world environments. However, these same characteristics can create or exacerbate product risks. We studied how industry developers (n=35) perceive, prioritize, and address the risks in their agentic AI products. We found that developers' perceptions of risk were closely tied to the qualities that made the product agentic, such as autonomy, tool use, and usage in a real-world context. Developers prioritized product and business risks before considering downstream societal risks like job displacement and end-user privacy. This prioritization also impacted developers' ability and motivation to mitigate agentic risks. Finally, developers lacked mature controls for containing agentic risks, often relying on constraining the same characteristics that make agents useful: e.g., autonomy and goal complexity. These findings reveal a capability vs. risk control tension in agentic AI development: developers need to address risks that emerge from agentic capabilities, yet they currently have limited support for doing so without constraining agentic functionality.

\end{abstract}


\section{Introduction}

Across academia, industry, and government, AI agents are framed as a new paradigm for software systems and technology governance because they can act autonomously with limited oversight, interact with and affect their environments, and operate across diverse roles, contexts, and tasks \cite{kasirzadeh_characterizing_2025, chan_harms_2023}. 
These same capabilities also create new and exacerbate known risks of generative AI systems: e.g., agents can violate policies or take unauthorized actions \cite{hart2026lobster}, misuse tools or environment resources that leak sensitive information or damage connected systems \cite{roth2026openclawskills}, and make actions harder to monitor and hold accountable \cite{hart2026kiro}.

Despite the growing excitement and concern around agentic AI, we know little about how developers who build user-facing agentic AI products perceive these emergent risks, decide which risks to prioritize, and implement controls to mitigate these risks. 
Understanding these points is essential because while emerging governance frameworks for agentic AI call for centering principles like meaningful human control, accountability, and monitoring \cite{imda2026_agentic_mgf,wef2025_ai_agents}, prior work has documented a substantial ``gap between principle and practice'' in developing human-centered and responsible AI \cite{winfield2018ethical,shneiderman2020bridging}.

To understand how developers perceive, prioritize, and mitigate agentic AI risks, we draw on the Security and Privacy Acceptance Framework (SPAF). 
SPAF was developed to explain end-users' adoption of security and privacy best practices \cite{das2022security}, and recent work has adapted its three barriers---\textit{awareness} of harms, \textit{motivation} to act, and \textit{ability} to address the harms---to study how \textit{practitioners} incorporate human-centered principles into AI product development \cite{lee2024idontknown}.
We use SPAF in a similar, developer-oriented way in the context of agentic AI product development.
Specifically, we ask three research questions:

\begin{itemize}
    \item[\textbf{RQ1}] To what extent does agentic AI shape developers' perceived risks of their products? (Awareness)
    \item[\textbf{RQ2}] How do developers prioritize addressing emergent agentic AI risks of their products? (Motivation)
    \item[\textbf{RQ3}] What constitutes addressing agentic AI risks for developers, and what affects their ability to do so? (Ability)
\end{itemize}

To answer these questions, we conducted semi-structured interviews with $N=35$ developers from different product teams at IBM. 
All participants had engaged in addressing agentic AI risks for their user-facing agentic AI products in some capacity. 
In line with the recent agentic AI literature, we define agentic AI products as products that employ foundation models (e.g., LLMs or large multimodal models) with access to tools (e.g., APIs, external services, computational resources), and have \textit{autonomy} to \textit{act in an environment} based on set goals, \textit{decide which actions to perform}, and \textit{execute multi-step actions to pursue goals} for end users \cite{pan_measuring_2026,kasirzadeh_characterizing_2025,wang_survey_2024,acharya_agentic_2025,miehling_agentic_2025}.


We found that developers focused more on risks that were directly proximate to product objectives for their AI agents (e.g., risks related to reliability and performance) and less on broader downstream risks (e.g., risks related to job displacement or privacy). 
Developers were particularly aware of risks related to agent policy violations, compounding failures, or unclear action accountability (\textbf{RQ1}).
Developers prioritized mitigating risks through an implicit model of business success: risks became more actionable when they threatened agent performance, user adoption, production safety, or client needs. 
Conversely, developers were more hesitant about mitigations that introduced opportunity costs, reduced product value, conflicted with development timelines, or fell outside of their perceived responsibilities (\textbf{RQ2}). 
Developers implemented a range of mitigation strategies and controls, such as human oversight, action constraints, access control, validation, and data sanitization.
Yet these controls often constrained the same properties that made the systems agentic in the first place: autonomy, adaptability, goal complexity, and environment complexity. 
Developers also lacked reliable assessment methods to evaluate whether these mitigations were effective (\textbf{RQ3}). 
In sum, our findings reveal a central tension in agentic AI product development: industry efforts to make products \textit{more agentic} create new risks, which developers then attempt to mitigate by making products \textit{less agentic}.

Our work makes the following contributions: 
\begin{itemize}
    \item We provide an empirical account of how developers perceive, prioritize, and mitigate risks entailed by agentic AI products. 
    \item We show how developers' \textit{risk awareness} is shaped by product-proximate agentic AI functionality, what \textit{motivates and inhibits} risk mitigation work, and what \textit{affects developers' ability} to implement risk controls.
    \item We extend the human-centered AI (HAI) literature on the principle-practice gap by showing how this gap manifests in agentic AI product development, where risk mitigation often requires developers to constrain the same agentic characteristics that make these products valuable.
\end{itemize}
\section{Related Work}
\paragraph{Agentic AI Risks}

Researchers have begun developing conceptual frameworks and governance approaches to systematically characterize and manage agentic AI risks, which are generally framed as extensions of existing generative AI risks \cite{Gan2026Navigating, Chang2026SystematicLit, Steenstra2025RiskOntology, li2026characterizingRisks, madkour2026_agentic_profile, bellogin2025ACM_Europe, Gangavarapu2025}. 
However, the frameworks differ in emphasis and scope, ranging from threat-actor taxonomies \cite{Gan2026Navigating}, to bias propagation arising from multi-stage workflows, \cite{condon2026Bias_Eval}, to domain-specific perspectives, risks to youth \cite{Chang2026SystematicLit}, or risks from psychotherapy agents \cite{Steenstra2025RiskOntology}. 
Governance-oriented approaches include mapping user-reported chatbot and agent risks to the NIST AI Risk Management Framework \cite{li2026characterizingRisks}, documenting agentic AI standards profile for practitioners and policymakers \cite{madkour2026_agentic_profile}, and developing recommendations on systemic agentic AI risks for the EU AI Act \cite{bellogin2025ACM_Europe}. Among these efforts, the AI Risk Atlas \cite{bagehorn_ai_2025} is distinctive in integrating risks from multiple frameworks into a unified knowledge graph that connects risk categories, governance frameworks, and mitigation strategies.

Although these studies provide valuable conceptualizations of agentic AI risk, they primarily reflect top-down perspectives derived from governance frameworks, expert analyses, and taxonomies. Instead, our work centers on practitioners' lived experiences in identifying, prioritizing, and mitigating agentic AI risks in practice.

\paragraph{Risk Mitigation Tools and Their Barriers}
Prior work in systems engineering, responsible AI, and security and privacy (S\&P) proposes mechanisms for building trustworthy AI systems by centering transparency and explainability \cite{thiebes2021trustworthy,floridi2018ai4people} and assigning accountability across the AI development life-cycle \cite{Li2023TrustworthyAI, Kenthapadi2023GenAI_and_RAI}. Organizational tools such as RACI (Responsible, Accountable, Consulted, and Informed) matrices specify ownership over training data, model tuning, deployment, and monitoring (e.g., \cite{glaser2024governance}), while documentation artifacts such as model cards, system cards, and datasheets for datasets, institutionalize transparency and traceability \cite{mitchell2019model, gebru2021datasheets}. 
Standards and regulations, including the EU AI Act, the NIST AI Risk Management Framework (AI RMF), and ISO/IEC standards (e.g., ISO/IEC 42001 for AI management systems), have codified many of these practices.

Yet studies of AI practitioners report confusion over which governance frameworks apply in a given context and difficulty integrating them into daily work \cite{lee2024idontknown}. Related work on S\&P practices identifies misconceptions, knowledge gaps, and implementation burdens that hinder adoption of recommended safeguards \cite{Li2018Coconut, Li2022Developer_challenges}. Lee et al. (\citeyear{lee2024idontknown}) applied the SPAF framework \cite{das2022security} to examine what prevents AI practitioners from adopting S\&P best practices in product development, showing how organizational and workflow constraints shape how developers prioritize and implement risk-mitigation measures.

We examine how developers operationalize governance practices when building \textit{agentic AI systems}. As recent work notes, increasing the capabilities of AI-embedded systems can expand the scope and severity of associated risks \cite{lee2024taxonomy, ehsan2026HCXAI}. This challenge is salient for agentic AI, which is being widely integrated and marketed into consumer products \cite{pan_measuring_2026,shome_why_2026}, and whose capabilities are expanded through tool use, external resources, memory, and autonomous action \cite{wang_survey_2024}.
We aim to better understand which agentic risks are emerging and prioritized in practice, and how developers seek to address them.
\begin{figure*}
\centering
\includegraphics[width=1.4\columnwidth]{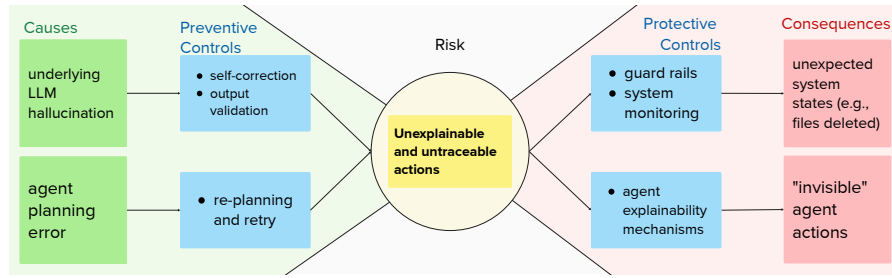}
\vspace{-2mm}
\caption{Example of a complete bow-tie analysis. For one agentic AI risk they had attempted to mitigate, participants identified causes, consequences, preventive controls for reducing causes, and protective controls for reducing consequences.}
\vspace{-2.5mm}
\label{fig:bow tie analysis}
\end{figure*}

\section{Method}
We conducted semi-structured interviews with 35 developers at IBM with direct experience building user-facing agentic AI products. 
This format let us ask consistent questions across participants while probing product-specific experiences perceiving, prioritizing, and addressing risks.


\paragraph{Pre-Study Questionnaire}
To ground each interview in participants' real product contexts, we first asked participants to complete a pre-study questionnaire covering three areas. 
First, participants described the user-facing agentic AI product they worked on. 
Second, they selected three human-centered AI (HAI) principles from a list that we adapted from prior studies \cite{sanderson_ai_2023,pant_ethics_2024} \footnote{The principles included \textit{Privacy Protection and Security}, \textit{Reliability and Safety}, \textit{Transparency and Explainability}, \textit{Fairness}, \textit{Performance and Testing}, \textit{Accountability}, \textit{Human-centered Values}, and \textit{Human, Social and Environmental Wellbeing}.} that they considered most concerning for their product.
Third, for each selected principle, participants rated the relevance of associated agentic AI risks to their product. We manually mapped agentic AI-specific risks identified in the AI Risk Atlas \cite{bagehorn_ai_2025} to each HAI principle.
The full list of agentic AI risks and their associated principles is provided in the Appendix.

The selected HAI principles helped prompt discussion of salient risks for \textbf{RQ1}. 
For \textbf{RQ2}, we randomly selected five risks each participant rated as relevant and used them to create a product-specific risk-ranking activity in Mural\footnote{\url{https://app.mural.co/}}; an example is shown in the Appendix.

\paragraph{Semi-structured Interviews}
All interviews were conducted remotely by the first author and lasted approximately 60 minutes. 
At the start of each session, we explained the study's purpose, obtained verbal consent for participation and recording, and informed participants that they could stop or withdraw consent at any time and that recordings would be de-identified.
Each interview began with participants introducing their agentic AI product. 
Then, building on and extending the SPAF \cite{das2022security}, we organized the interview around our three research questions on how agentic AI shaped developers' awareness (RQ1), motivation (RQ2), and ability (RQ3) to address product risks. 


To address \textbf{RQ1}, we used participants' three HAI principles from the pre-study questionnaire as discussion prompts.
For each principle, we asked participants to recall a recent product discussion about a related risk: how they defined and scoped it, and how it related to their product's development.

To address \textbf{RQ2}, participants ranked five product-relevant agentic AI risks selected from their questionnaire responses using a Mural-based ranking board.
We asked participants to explain their ranking and why some risks were more or less important to address.
We analyzed these responses to identify factors shaping risk prioritization in user-facing agentic AI product development.

To address \textbf{RQ3}, participants completed a bow-tie analysis in Mural for one agentic AI risk their team had attempted to mitigate \cite{koessler_risk_2023} (Figure~\ref{fig:bow tie analysis}). 
Participants first placed the risk at the center of the diagram. 
They then identified potential \textit{causes} of the risk (e.g., agent errors, tool failures, unexpected user behavior), followed by \textit{preventive controls} intended to address those causes.
Next, they identified possible \textit{consequences} (e.g., incorrect outputs, security incidents, loss of user trust), followed by \textit{protective controls} intended to reduce those consequences. 
In our analysis, we focused on participants' controls: the actions, tools, artifacts, and processes they used to mitigate the causes or consequences of agentic AI risks, as well as the challenges involved in applying those controls.
This allowed us to examine how developers operationalized risk mitigation in practice.
The full interview protocol is included in the Appendix.

\paragraph{Recruitment and Participants}
We recruited IBM developers with experience building and deploying user-facing agentic AI products and addressing risks in that process.
At the time of the study, IBM maintained an internal AI Ethics Board and enterprise AI governance tooling, though we did not directly assess how well these structures covered agentic AI-specific risks.
This study was reviewed by an internal committee, and this study received approval subject to restrictions on demographic data collection. We were not permitted to collect participants' age, gender identity, or personally identifiable information. Participants were compensated with a gift card equivalent to \$25 USD.

In total, we recruited 35 developers from different product teams across the company, working on agentic AI products across a range of domains\footnote{Participant information is included in the Appendix.} --- most commonly Customer Support (n=7), Information Technology (n=7), and Corporate Services (n=6).

\paragraph{Data Analysis}
All interviews were audio-recorded and transcribed. 
We conducted an iterative open-coding process \cite{corbin2008basics} aligned with our three research questions. 
The first author coded ten interview transcripts and developed a preliminary codebook through discussion with two other authors. 
Two additional authors then joined the coding process and were trained on the codebook.
The three coders divided the remaining transcripts and coded them individually, meeting regularly to discuss codes and emerging themes. 
They also reviewed one another's coded excerpts to promote consistency and resolved disagreements through discussion. 
The three coders were employees of IBM at the time of the study. To guard against affiliation bias, codebook development and thematic refinement were conducted in discussion with the non-affiliated authors throughout analysis.

\begin{figure}
\centering
\includegraphics[
  width=\columnwidth,
]{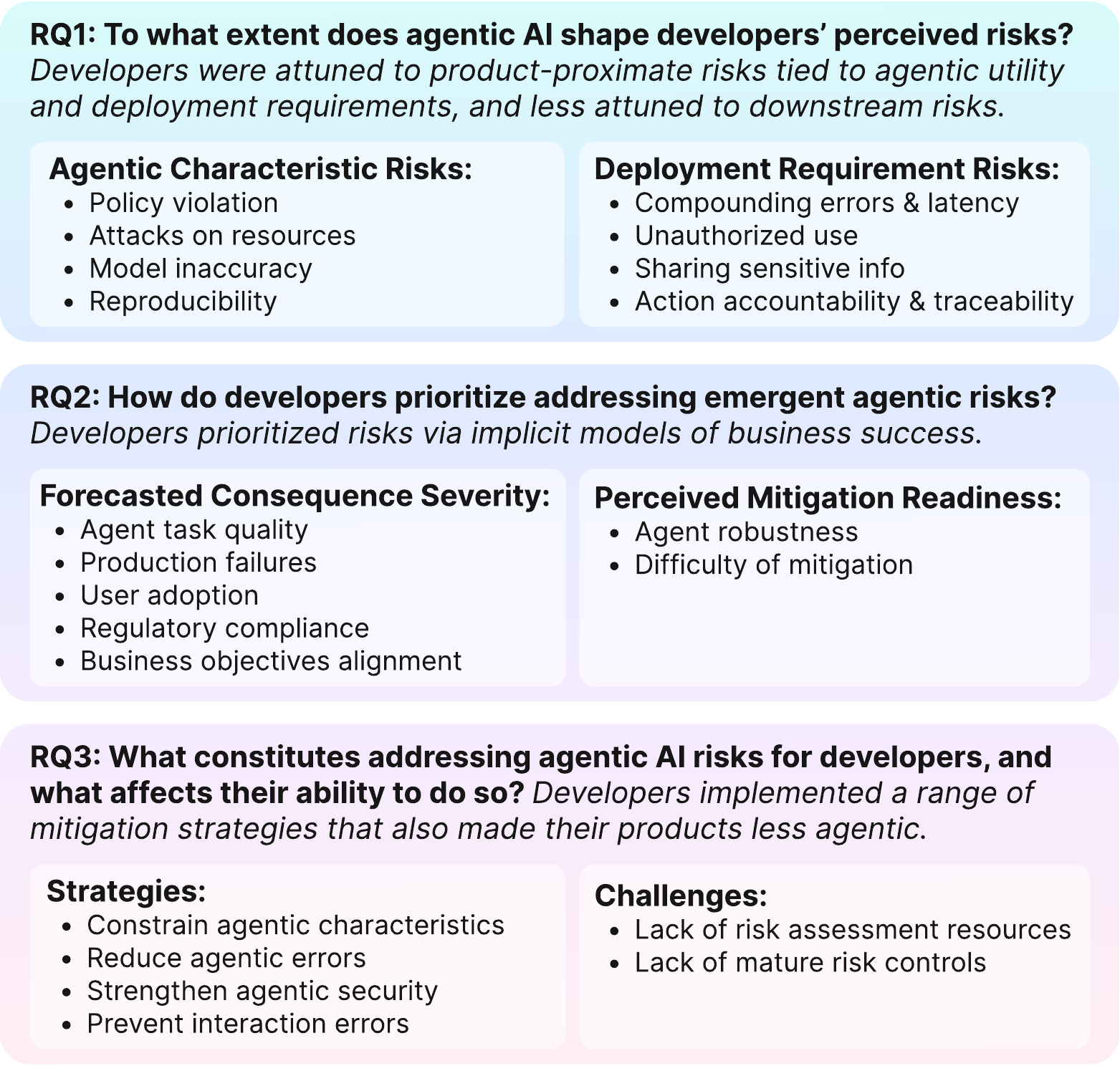}
\vspace{-3mm}
\caption{We answer our research questions by showing how agentic AI shaped developers' awareness (RQ1), motivation (RQ2), and ability (RQ3) to address product risks.}
\vspace{-1mm}
\label{fig:result-summary}
\end{figure}

We organized our findings around the three SPAF-grounded research questions \cite{das2022security} (Figure \ref{fig:result-summary}) and report how frequently participants discussed each theme.

\section{RQ1: How Agentic AI Shapes Developers' Perceptions of Product Risks}

Following the SPAF \cite{das2022security}, we first examined the extent to which agentic AI shapes risk \textit{awareness}.
We first analyzed what participants considered to be the most concerning risks in developing user-facing agentic AI products.\footnote{Using the AI Risk Atlas \cite{bagehorn_ai_2025} as a baseline to categorize participants' concerns, we identified ten agentic AI risks in total, three of which are not fully captured by the atlas: agent policy violation, compounding errors, and compounding latency.} 
We then examined whether and how these risks were shaped by the agentic nature of their product.
 
We found that participants were generally attuned to risks tied to two dimensions: \emph{agentic AI characteristics} and the \emph{requirements of real-world deployment} (Figure \ref{fig:result-summary}).
At the same time, developers' awareness remained bounded by individual judgment, prior experience with non-agentic AI systems, and locally available organizational knowledge.
Thus, they were less attuned to broader, downstream risks that are less proximate to agent utility or deployment.

\subsection{Agentic AI Characteristics Led to New Risks} 
Developers shared that the characteristics of agentic AI---autonomy, environment complexity, goal complexity, and adaptability---either introduced agentic-specific risks or created agentic-specific instantiations of known AI risks.

\paragraph{Agent autonomy enabled agent policy violation risks.} 
Developers built AI agents that achieve goals with limited human intervention or supervision \cite{kasirzadeh_characterizing_2025,acharya_agentic_2025}, typically relying on prompt-based ``agent policies'' to define what agents should or should not do, such as enforcing user permissions or restricting information sharing with third-party tools.
Yet the same autonomy that made agents useful also raised concerns about failure to comply with predefined action policies.
One key risk participants flagged was \textbf{agent policy violation (8/35)}, driven by limited oversight over agents' reasoning and actions, especially when agents could bypass access policies or other organizational constraints. 
For example, P26 described the central risk of their agent as the possibility that \textit{``[agentic] rules can be violated... So the constraints [to the agent] are not really constraints.''}

\paragraph{Environment complexity exposed agent resources to attacks.}
Developers built AI agents that can act in complex and connected environments to achieve their goals \cite{kasirzadeh_characterizing_2025,acharya_agentic_2025}.
Agentic AI products are integrated with internal or third-party APIs, platforms, tools, and data sources.
As a result, \textbf{agents introduced new attack vectors into environmental resources (10/35)}. 
Developers worried that agents could misuse external tools or perform harmful actions that unintentionally modify or damage resources in their environment. 
P2's IT agent, for instance, was flagged by the cybersecurity team for \textit{``trying to find and delete root folder.''}

\paragraph{Agent goal complexity exacerbated model accuracy risks.} 
Developers also built AI agents to achieve a wide range of goals \cite{kasirzadeh_characterizing_2025,acharya_agentic_2025}. 
As participants assigned agents more complex tasks, they found it harder to ensure reliable task success. 
Because all participants' agents were powered by generative AI, participants described how agents exacerbated issues with \textbf{model inaccuracy (19/35)}: errors compound over multi-turn autonomous execution. 
For example, P35 set a high minimum performance threshold for their IT agent to reduce errors cascading across turns: \textit{``if it goes below 80\%, that means the agent is actively harming your system rather than helping your system.''}

\paragraph{Agent adaptability created new instantiations of reproducibility risks.}
Finally, developers built AI agents that adapt and react to novel or unexpected circumstances \cite{kasirzadeh_characterizing_2025,acharya_agentic_2025}.
Participants valued agentic AI as an adaptable solution that could handle both expected and unexpected use cases. However, this adaptability introduced \textbf{reproducibility (8/35)} risks --- the difficulty of replicating agent behavior or output --- 
stemming from both model variability and the agent's ability to select different tools, actions, and plans across similar situations. 
For example, P7 shifted from a structured keyword matching retrieval process to an agentic solution that constructed Jira Query Language from user inquiries \textit{``to make a more adaptable solution,''}, but noted that this came \textit{``at a cost of repeatability.''} 


\subsection{Agentic AI Requirements Increased Risks Related to Latency, Security, and Monitoring}
We also found that the requirements of developing and deploying agentic AI systems exacerbated concerns about performance, latency, security, accountability, and traceability.

\paragraph{Layered agent architectures exacerbated compounding errors and latency.} 
Participants often described their agentic systems as complex architectures composed of multiple models, tools, agents, and orchestration layers. This complexity led developers to identify \textbf{compounding errors (7/35)}, wherein failures accumulate across agentic workflow components, requiring additional debugging or monitoring. 
Although errors in generative AI-powered systems are not new, participants perceived agentic architectures as increasing their intensity and frequency. 
As P25, who built a multi-agent system, put it, they: \textit{``have multiple agents, so multiple points of failure.''}
These layered architectures also exacerbated \textbf{compounding latency (4/35)} risks for some participants, as each added tool, retrieval step, or orchestration layer could increase response time.

\paragraph{Agentic AI introduced and exacerbated security threats.} 
Participants perceived that agentic AI introduces new security threats, including \textbf{unauthorized use (3/35)}, which concerns attackers gaining access to the AI agent or its components. 
During testing, for instance, P2 found that a \textit{``malicious party can hack into that agent and take complete control over the system where the agent was being executed.''} 
Others identified risks of \textbf{sharing IP/PI information (14/35)}, which concerns agents leaking intellectual property or personally identifiable information from their systems, primarily company repositories or customer data, via tool use or interactions with the agent environment. 
Agentic AI also exacerbated known generative AI security risks because those risks could apply across agentic system components and tools. For example, P29, whose agent handled bank transactions, emphasized the need to protect ``every step'' of the agent from prompt injection attacks.

\paragraph{Agent deployment necessitates accountability, explainability, and traceability risks.}
Deploying agents in real environments required developers to consider the \textbf{unclear accountability of agent actions (8/35)}, where responsibility for an agentic AI system's actions becomes difficult to assign. 
In P31's agent, \textit{``a multi-agent system built by different teams and aggregated,''} making it difficult to pinpoint responsibility when an agent action fails. 
Participants also differed in how they assigned responsibility between users and agentic systems. P26 framed the agent as \textit{``just a tool for the office worker,''} suggesting\textit{``if the tool is not working well, the office worker is to blame,''} while others like P32 stressed distinguishing agent failures from end-user responsibility: \textit{``if the agent does something wrong, you don't want to fire an employee because of it.''}

Finally, participants raised concerns about \textbf{unexplainable and untraceable agent actions (16/35)} being a critical barrier to end-users' trust and comfort delegating actions --- where explanations, lineage, or source attribution are difficult, imprecise, or unavailable. 
P16 explained that clients deploying agents in high-stakes or regulated environments needed records of the agent's \textit{``each and every step''}, understanding how an agent arrived at its outputs on what data.


\subsection{Awareness Barriers: Novelty and Hype Hindered Developers' Awareness of Agentic AI Risks}

While developers were sensitized to the risks introduced and exacerbated by the design, development, and deployment of their novel agentic systems, we found that their risk awareness was also inhibited by limited organizational guidance and uncertainty around the evolving agentic AI landscape.

\paragraph{Developers relied on individual judgment to identify agentic AI risks in lieu of specific guidance.}
Participants (9/35) described agentic AI as too new for organizations to have established practices, resources, expertise, or clear responsibility structures, leaving developers to \textit{``learn on your own,''} with \textit{`zero resources available''} (P30), extrapolate from prior non-agentic AI experiences, and make situated judgments about which harms mattered for their products. As P7 noted, a core challenge in identifying risks was \textit{``the lack of the availability of the seniors... There are not many people building AI agents... there's no one ready to take the responsibility.''}


\paragraph{The hype around agentic AI made it difficult to clarify the risk landscape.} 
For some product teams (8/35), agentic AI remained speculative and contested, requiring developers to negotiate what agents could realistically do, what constituted a useful agentic product, and what risks were worth addressing.
P9 described this as \textit{``trying to dispel what is hype and what is actually useful.''} At the same time, developers also had to navigate a rapidly expanding ecosystem of testing frameworks and evaluation tools for agentic systems, requiring developers to \textit{``figure out the noise from the quality [ones]''} (P12) when assessing the real risks.

\paragraph{\textit{RQ1 Summary:}} Taken together, our findings show that developers' risk perceptions were shaped by risks newly created by agentic AI characteristics and by risks created or exacerbated through agentic AI product development and deployment. However, this risk awareness was uneven: developers were better at identifying risks directly tied to their agentic AI product's envisioned utility and deployment, while broader, downstream societal harms were harder to anticipate due to limited guidance and the evolving agentic AI landscape. We revisit developers' potential agentic AI risk-awareness blind spots in the Discussion section.

\section{RQ2: How Practitioners Prioritize Addressing Emergent Agentic AI Risks in Their Products}

Grounded in the \textit{motivation} layer outlined in the SPAF \cite{das2022security}, we next analyzed how participants described the way they (de)prioritized agentic AI risks for their products during the risk-ranking activity --- i.e., what makes a risk ranked higher/lower than the others.
Our participants gave higher priority to risks with greater \textit{forecasted consequence severity}. 
Their prioritization also considered their \textit{perceived readiness to mitigate} those risks (Figure \ref{fig:result-summary}).

Beyond the risk-ranking activity, our interviews revealed barriers that inhibited developers from prioritizing risk mitigation during agentic AI development. 
These barriers were primarily organizational: risk mitigation often competed with other product goals, introduced opportunity costs, and became difficult to prioritize when ownership was unclear. 

\subsection{Developers Prioritized Product-Proximate Risks; De-prioritized Broader Risks}
Throughout the risk-ranking activity, our participants frequently raised five consequence-oriented considerations that shaped their risk prioritization: agent task quality, production failures, user adoption, regulatory compliance, and alignment with business objectives.

\paragraph{Agent task quality.}
Risks that could undermine the quality of an agent's task performance, including its accuracy, consistency, and efficiency, were a priority for many participants (14/35). 
For accuracy, developers assessed whether a risk would prevent the agent from completing its intended function, as P1 explained, \textit{``I view [risks] more as the pieces that would inhibit function.''}
Participants also prioritized risks that produced inconsistent outcomes across iterations, threatening reliability.
For example, P12 prioritized the traceability of agent actions because if the product team \textit{``can't reproduce the output, it's really hard to fix whatever comes out.''}
For efficiency, developers considered whether a risk would make their agent resource-intensive, such as making it \textit{``too slow''} (P31) or \textit{``incurring more cost''} (P4).

\paragraph{Production failures.} 
Some participants (12/35) prioritized risks based on the severity of failures that could occur once the agent was deployed in production. 
These assessments were context-dependent. 
For example, P10, whose agent handled credit card information inquiries, emphasized that hallucinations could not be treated as \textit{``just a silly mistake''} because \textit{``these are regulations. These could cost the users heavily, so mistakes are not tolerable for this product.''} 
Participants also assessed production attack surfaces. 
As P18 noted, a \textit{``malicious prompt''} could allow someone to \textit{``play with the resources or the tools or tamper with or poison the tools that the agent has access to.''}

\paragraph{User adoption.}
Some participants (13/35) prioritized risks based on their consequences for user adoption, especially user trust and usability. 
Developers worried that inconsistent or opaque agent behavior could erode users' willingness to use their systems. 
At the same time, participants did not necessarily define trust as requiring a risk-free system; rather, they emphasized that users should understand risks to calibrate their trust. 
As P15 put it, \textit{``if it's transparent, people can take the risk, and people can avoid the risk and mitigate risk. Transparency is a must.''} 
Participants also prioritized risks that harmed usability. 
P1, for instance, prioritized addressing redundant agent actions because repeated reflection, tool invocation, and duplicate steps could \textit{``bring frustration to the user and make it unusable.''}

\paragraph{Regulatory compliance.}
Participants (11/35) assessed severity through the consequences of non-compliance, 
prioritizing risks that could produce financial or legal repercussions, especially in regulated contexts. 
Consistent with prior work on developer motivations around responsible AI and privacy \cite{tahaei2021privacy,lee2024idontknown}, external regulatory mandates such as the GDPR and the EU AI Act served as important catalysts for risk prioritization.

\paragraph{Alignment with business objectives.}
Finally, some participants (9/35) prioritized risks based on whether they aligned with client concerns or current business priorities; 
P21 prioritized personal data exposure to external tools because \textit{``clients are always concerned about sharing the personal info like ID number [with external tools].''} 
Conversely, other developers de-prioritized risks that were less urgent for current business goals --- such as how P2 de-prioritized the risk of exploiting users' trust in their AI agent because it was \textit{``much longer along the road map, not right now.''}

\subsection{Developers Prioritized Risks That Were Harder to Measure and Mitigate}
Participants also prioritized risks by assessing whether their current or near-term systems were ready to mitigate them. 
These judgments depended on two factors: perceived agent robustness and perceived difficulty of mitigation.

\paragraph{Perceived agent robustness.}
Participants (14/35) assessed mitigation readiness through their confidence in the robustness of the agentic system stack, including the underlying model, tools, resources, and architecture. 
When participants trusted these components based on available evidence, they tended to de-prioritize the associated risk.
P12, for instance, treated function-calling hallucination as less concerning because their team had applied the ReAct framework \cite{yao2022react} and because their agents \textit{``don't hallucinate anymore, especially with the larger models.''} 
Participants (13/35) also assessed robustness through existing data protection mechanisms and data flows. 
For example, an existing verification mechanism could decrease risk (P1), while dependencies on fragile cloud services or external systems could increase it (P30).

\paragraph{Perceived difficulty of mitigation.}
When a risk had not yet been mitigated, participants assessed how difficult remediation would be in practice, with many (9/35) considering whether available off-the-shelf solutions were available and whether mitigation could be implemented with minimal disruption to product development.
Counterintuitively, poorly understood mitigations drove higher risk prioritization: P32 prioritized proprietary data leakage because \textit{``there are more nondeterministic ways to make the agent leak information, and the way to mitigate that is not fully understood yet.''} 
Conversely, risks were de-prioritized when the required mitigation appeared straightforward or when implementing it involved limited integration work.


\subsection{Motivation Barriers: Organizational Inhibitors Hindered Developers' Risk Mitigation Work}
Our findings suggest that developers prioritized mitigating risks proximate to their product success.
Simultaneously, our interviews show that such product-success-driven approaches also introduced organizational inhibitors that constrain developers from prioritizing risk mitigation work, including opportunity costs and unclear ownership. 

\paragraph{Opportunity costs and trade-offs.}
Most participants noted that risk mitigation involves significant opportunity costs and trade-offs: with tight schedules and limited resources, justifying mitigation was complicated by abstract benefits against tangible costs. We observed this resulting de-prioritization of risk mitigation in several forms. 

Developers described trade-offs between mitigation and \textbf{agent performance (12/35)}: 
as we will note in RQ3, many mitigation strategies improved safety by constraining agent behavior, such as restricting tools, limiting data access, or adding guardrails. 
Yet these same constraints could also reduce agent value, flexibility, or accuracy. 
As P3 put it, \textit{``the more you limit these tools, the less valued [it] is,''} framing the tension as \textit{``how far do we swing to give the user the ability to do things but also not shoot themselves in the foot?''} 
Some participants also noted that mitigation reduced \textbf{cost-efficiency (6/35)}, as additional checks required \textit{``extra computation''} (P18), increasing both time and monetary cost. 


Some participants described risk mitigation as competing with \textbf{business priorities (10/35)} and product timelines. 
As P32 noted, limited engineering capacity meant that the team \textit{``prioritized other functionalities first.''} 
Others deferred mitigation because of \textbf{time pressure (9/35)}, especially when products or models changed faster than teams could evaluate and integrate new safeguards. 

Many developers also described mitigation as increasing \textbf{engineering costs (22/35)}. 
Controls added complexity to already complex agentic AI systems, including guardrails for prompt templates, tool-access policies, communication protocols, system state, and memory. 
These costs continued after deployment because controls had to be maintained as risks evolved (e.g., new prompt injections). 

\paragraph{Risk ownership externalization and misalignment.}
Building AI agents was often a cross-team effort, and participants (15/35) described limited ownership over risks tied to components built by external teams or falling outside their perceived responsibility. 
P10 drew a clear boundary around their role in agentic AI development: \textit{``I'm the tech guy. I can only make sure it's as reproducible as possible, but when it's not, it's basically the organization's responsibility,''} dismissing regulatory concerns as \textit{``not really my concern.''} 
This externalization of mitigation responsibility echoes prior responsible AI work showing that practitioners often distribute ethical responsibility across sociotechnical networks, leaving accountability for risk mitigation unclear \cite{orr2020attributions,rakova2021where,lee2024idontknown}.

Some participants (6/35) also described misaligned priorities across teams. 
For example, P8's agent depended on a tool built by another team whose codebase introduced a vulnerability that was \textit{``not a priority for them,''} forcing P8's team to seek workarounds because they lacked direct ownership over the source of the risk.

\paragraph{\textit{RQ2 Summary:}} Overall, developers' prioritization of risk mitigation was motivated by both risk-level and product-level factors: at the risk level, developers prioritized risks that threatened product performance, deployment, compliance, or adoption, especially when existing systems were not ready to mitigate them; at the product level, mitigation competed with other objectives and became harder to prioritize when ownership was unclear or cross-team priorities were misaligned. In short, unless risks could be tied to business success, developers often encountered more inhibitors than motivators when prioritizing agentic AI risk mitigation.

\section{RQ3: What Constitutes Agentic AI Developers' Risk Mitigation Approaches}

Finally, as outlined in the SPAF \cite{das2022security}, we examined developers' \textit{ability} to translate intentions into risk mitigation.
Accordingly, using the bow-tie analysis framework (Figure \ref{fig:bow tie analysis}), we analyzed the \textit{controls} participants used to address the causes or mitigate the consequences of agentic AI risks. 
We found that developers addressed agentic AI risks through five mitigation strategies: constrain agentic characteristics, reduce agentic errors, strengthen security for agent-mediated data and tool use, prevent human-agent interaction errors, and apply best practices from LLM and software engineering. 
Across these strategies, a recurring pattern emerged: mitigating agentic AI risks often required developers to limit the very characteristics that made their systems agentic, including autonomy, adaptability, tool use, and open-ended goal pursuit. 
Developers' ability to implement these controls was also limited by immature assessment methods and by the lack of established, reliable controls for agentic AI systems (Figure \ref{fig:result-summary}).

\subsection{Mitigation Strategies for Agentic AI Risks}

\paragraph{Constrain agentic characteristics.}
The first set of risk mitigation strategies limited agentic characteristics directly, i.e., autonomy, environment complexity, goal complexity, and adaptability. However, these controls introduced trade-offs that narrowed the agent's scope of action, reduced its flexibility, or inserted additional human oversight.

To \textbf{constrain agents' autonomy (7/35)}, participants implemented human-in-the-loop workflows requiring users to confirm, approve, or revise agent actions, which were especially vital in high-stakes operations. 
P18's agent, for example, requested confirmation before executing actions like \textit{``run the delete policy function.''} 
Similarly, P22 used human-in-the-loop review, allowing users to \textit{``modify the plan''} when they were unsatisfied with customer-support plans.

To \textbf{constrain agents' environment complexity (8/35)}, participants limited what tools agents could access and how those tools could be used. 
One common approach was \textit{tool specification}: defining the purpose and scope of each tool so that agents had fewer ambiguous options. 
As P3 explained, \textit{``our tools that we've given the agent access to are very specific,''} with little ambiguity about what each tool can do. 
Developers also made agents sensitive to the availability and readiness of tools that interact with the environment. 
For instance, P1 added a planner layer that temporarily adjusted the agent's plan based on which tools were available.
Some participants further reduced risks from environment complexity by designing agent actions to be reversible, ensuring that \textit{``whatever the agent does can be undone''} (P35).

To \textbf{constrain agents' goal complexity (5/35)}, participants recalibrated what agents were expected to do based on the capabilities of the underlying model. 
Rather than allowing agents to pursue broad or difficult objectives, developers increased task success by \textit{``limiting the problem space''} (P9) or by constraining the agent to \textit{``easier problems''} (P4). 
For longer multi-turn tasks, some developers added interim human verification, so users could steer the agent before it diverged too far from the intended path. 
As P9 explained, this prevented users from having to \textit{``wait for an hour before getting any feedback''} and allowed them to provide input while the agent was still acting.

To \textbf{constrain agents' adaptability (15/35)}, participants introduced predetermined actions, responses, and output constraints to make agent behavior more predictable and reproducible. 
For example, P7 described that their agent was now implemented as a ``predefined workflow'' rather than a fully adaptable agentic solution.  
P30 similarly hard-coded parts of the agent to steer its recommendations, while others \textit{``limit[ed] the response format''} (P26) or implemented a topology graph to keep their agent \textit{``confined to the problem space''} (P9).

\paragraph{Reducing and recovering from agentic errors.}
A second set of risk mitigation strategies focused on reducing or recovering from agentic errors, including reasoning errors, incorrect planning or execution, budget overruns, and tool use errors \cite{agashe2025agent, crispino2023agent, winston2025taxonomy}.
Participants used these controls to prevent incorrect outputs, reduce inefficiency, and avoid wasting time or computational resources.

First, developers implemented \textbf{action timeouts (3/35)} when agents failed to complete tasks within expected time or budget limits. 
For example, P6 explained that if the agent continued for too many iterations \textit{``we will be stopping it there,''} while P1 configured the agent to treat delayed tool responses as a non-response once a timeout was reached.

Second, developers implemented \textbf{action validation (14/35)} to check whether agents completed tasks correctly. 
Participants added model-based validation layers, such as validator agents or LLM-as-a-judge mechanisms, to assess an agent's output, reasoning, tool use, or intermediate results. 
P7, for example, implemented a validator agent to check whether the agent's explanation aligned with its recommended steps.
Others, like P8, validated tool outputs before passing them to downstream services, particularly when agents drew information from multiple sources. 

Third, developers implemented \textbf{self-improvement mechanisms (5/35)} that allowed agents to reflect, retry, or improve their responses and actions. 
For more complex tasks, retries could also substantially improve task completion. 
P9, for instance, described a remediation loop that allowed their IT agent to check whether alerts had cleared and continue trying if they had not, improving ``remediation'' performance even when the agent's initial diagnosis remained unchanged.

\paragraph{Strengthening security for agent-mediated data and tool use.}
A third set of mitigation strategies adapted established security practices, such as access control and input/output sanitization, to agentic workflows. 
Developers adapted these practices not only by protecting a model's prompt or response, but also by constraining how agents acted across user permissions, tools, and data sources.

Participants implemented \textbf{agentic access control (9/35)} by governing both what data agents could access and which tools they could invoke. 
Some aligned the agent's data permissions with those of the user (e.g., role-based authentication), ensuring the agent could not retrieve data on behalf of unauthorized users.  
Others controlled tool combinations to prevent sensitive data from flowing into inappropriate services.
P20 explained that if a sales database contained confidential data and Google Search was not approved to receive it, the agent would \textit{``never call those two combinations.''}

Participants also used \textbf{data-flow sanitization (11/35)} to filter agent inputs and outputs. 
These controls served as boundary checks on what information could enter or leave downstream agent tool calls. 
For inputs, participants described safety checks for uploaded files, filtering harmful user instructions, protection of system prompts, and substitution of personal information with dummy data. 
For outputs, developers masked sensitive information, while also accounting for task-specific needs. As P29 noted, banking users may legitimately need to see sensitive details like a ``transaction ID'' to complete a task.

\paragraph{Preventing human-agent interaction errors.}
The fourth set of strategies targeted errors arising from human-agent interaction, focusing on helping users form accurate mental models of agent capabilities and limitations, reducing instruction-related errors, and preserving user trust.

To this end, some developers (13/35) made agent behavior more visible through traces, notices, explanations, or source references. 
P29, for example, described traceability as a way to show users what information the agent relied on, as \textit{``traceability is our mitigation to not lose trust, [and] to gain trust.''} 
Participants also viewed source references as a way to communicate AI agents' limitations more broadly: P26 noted that visible citations help users understand that \textit{``AI can make mistakes''} and enable them to verify whether the agent summarized information correctly.

Other developers (6/35) reduced human-agent interaction errors by helping users specify goals and tool-relevant intent. 
Rather than treating user input as a prompt for a single response, these controls helped translate vague requests into plans that agents could follow, route to appropriate tools, and execute. 
P1 provided query examples and documentation to help users understand which inputs would invoke the right tools. 
Others added query-rewrite layers: P18 explained how a vague request like \textit{``write a policy to protect my data''} could be rewritten into a more actionable sequence of steps that helped the agent \textit{``do a much better job.''}

\paragraph{Applying best practices from LLM and software engineering.}
Participants also drew on established LLM and software engineering best practices to improve agentic systems' reliability. 
For LLM-specific practices, participants (9/35) tested and selected underlying models and parameters, improved the quality of agent data sources, and curated real-world task data for training or fine-tuning rather than relying on prompt engineering alone. 
For software engineering practices, participants (14/35) implemented redundancy and backup mechanisms, monitored tool and agent behavior, conducted code review and testing, and relied on reputable or established tools for agent use. 
While these practices were not specific to agentic AI, developers saw them as critical to bridging the gap between model capabilities and real-world agentic deployment contexts.

\subsection{Ability Barriers: Immature Tooling Hindered Developers' Ability to Mitigate Agentic AI Risks}

While participants implemented a wide range of controls, they still faced ability barriers that limited their capacity to address agentic AI risks. 
Developers lacked mature controls, reliable ways to assess agentic AI risks before choosing controls, and methods to validate the effectiveness of selected controls at mitigating risk post-deployment.


\paragraph{Lack of risk assessment resources.}
Participants (10/35) struggled with agentic risk assessment, which was often a prerequisite for selecting and evaluating mitigation strategies --- 
particularly with defining and collecting ground truth data, adapting general benchmarks to domain-specific agentic use cases, and ensuring assessments reflected real-world agent behavior. 
P15, who worked on a legal and compliance agent, noted \textit{``there is almost no way to say that this is correct, this is incorrect.''} 
P24, on the other hand, explained that their agent was so domain-specific that \textit{``you should have your own examples to test on.''} Yet it was difficult to obtain enough examples from their client for effective risk evaluation.
Even when participants could use established benchmarks, they worried that these benchmarks might not capture real deployment conditions. 
As P2 explained, \textit{``sometimes unexpected things happen in a live environment,''} and controlled assessments may not capture how an agent adapts to such noise. 
Thus, assessment challenges limited developers' ability to know which controls were needed and whether implemented controls were sufficient.

\paragraph{Lack of mature risk controls.}
Our participants (13/35) also struggled because agentic AI was too new for mature controls to exist. 
Some participants described agent steering and response control as \textit{``a very open area of research''} rather than a solved engineering practice, even though it is something developers \textit{``desperately need''} (P3).

Where usable controls or best practices --- e.g., data-flow sanitization, validator agents, action reversibility --- did exist, participants found them limited in \textit{scope}, \textit{effectiveness}, and \textit{consistency}. 
Some controls were not applicable across all agentic contexts: developers could not always prepare backup endpoints for every tool (P8), or not all agent actions were reversible (P31, P35). 
This case-specificity meant that the same control strategy could work for one agent, tool, or environment but fail to cover another.

Participants (6/35) also acknowledged that existing controls often reduced risks without fully mitigating them. 
For example, P7 noted that their validator agent was still under-trained with domain-specific data and could not always determine whether the agent's explanation was relevant to the steps it recommended. 
Others were very candid about the weakness of available controls. 
Describing their use of a disclaimer on potential wrongdoings as a mitigation strategy, P31 said: \textit{``it's a crappy solution. It's just giving up. It's easy to put a disclaimer.''}

Finally, participants (17/35) worried that controls were themselves inconsistent. 
Many controls, such as data-flow sanitization, validator agents, and LLM-as-a-judge mechanisms, were also powered by LLMs or agentic components. 
As a result, the control inherited some of the same nondeterminism it was supposed to mitigate. 
P25 summarized this problem: \textit{``you are introducing another point of failure because you're introducing another stochastic approach.''} 
Similarly, when access control or safety rules were implemented as agent policy, developers worried that agents could still violate those policies, echoing the agent policy violation risks described in RQ1.

\paragraph{\textit{RQ3 Summary:}} Overall, our findings show that developers addressed agentic AI risks through a wide range of controls, such as constraining agentic characteristics, reducing and recovering from agentic errors, strengthening agent security, preventing human-agent interaction errors, and applying conventional engineering best practices. Yet a central tension emerged in agentic AI risk mitigation: mitigating risks often meant limiting the same characteristics that made a product agentic, i.e., autonomy, adaptability, tool use, and open-endedness. Moreover, their ability to mitigate agentic risks was constrained by assessment methods and controls that are limited in scope, effectiveness, and consistency.

\section{Discussion}
We synthesize promising avenues to better enhance developers' \textit{awareness} of, \textit{motivation} to act on, and \textit{ability} to address agentic AI risks when building user-facing agentic AI products.

\subsection{Improving Awareness of Agentic AI Risks}


Our RQ1 findings show that developers were more aware of risks closely tied to agent utility and less attentive to risks less directly connected to product deliverables. 
This product-proximate awareness can create blind spots. 
For example, a survey of around 10,000 participants across ten countries found that ``job availability'' and ``privacy'' were among the public's top concerns about AI \cite{kelley_there_2023}. 
These concerns may intensify as AI agents become more autonomous and embedded in everyday work: agents may automate multi-step workflows that previously required human labor, while also requiring access to users' interaction logs, personal data, and third-party tools to act on their behalf.
Yet our participants did not foreground these threats as central concerns. 
This mismatch does not mean developers were unaware of these harms, but suggests that available risk-identification structures were organized around product utility, adoption, and deployment.
Downstream societal harms beyond any one product's direct purview, like privacy and job availability, had few organizational entry points to surface.

We envision two approaches for broadening developers' risk awareness: agentic AI risk scanning and simulated attacks.
First, recent work in responsible AI has used generative AI to help practitioners foresee potential downstream harms during product ideation and prototyping \cite{lee2026privy,wang_farsight_2024,buccinca2023aha}. 
Future work can adapt these risk-scanning tools to agentic AI by foregrounding agentic characteristics and data flows, surfacing harms that lack a direct product-utility link. 
Second, red teaming is a common practice in privacy and security for simulating how adversaries might compromise a product \cite{das2022security}, and it has already been adapted in AI contexts to reduce harms \cite{ganguli2022red}. 
Recent efforts have also begun to leverage LLMs and LLM-based agents to automate and scale adversarial testing across realistic tool-use, prompt-injection, and penetration-testing scenarios \cite{ruan2024identifying,debenedetti2024agentdojo,deng2024pentestgpt}. 
Future work can explore how LLM-powered red teaming might help developers surface and track downstream harms as agentic characteristics evolve.

\subsection{Improving Motivation to Address Agentic AI Risks}



Our RQ2 findings suggest that developers were motivated to address risks tied to product and business success, such as agent performance, but less so for risks that competed with product goals, increased costs, or that were perceived to fall out of their responsibility.
We envision two approaches for strengthening motivation to engage in agentic AI risk mitigation: pro-social and empathy-based design.

For \emph{pro-social design}, organizations could develop shared repositories of agentic AI risk mitigation ``light patterns'' with before-and-after examples of how product designs and outcomes changed.
Such repositories could provide social proof for HAI-aligned agentic AI development.
For \emph{empathy-based design}, recent work has explored narrative- and empathy-based approaches for helping practitioners translate abstract risks into relatable human consequences \cite{chen2024an,nahar2026i}. 
When incorporated into developer education and product development life-cycles, empathy-based design could help developers better understand how agentic AI risks affect users and other stakeholders. 
Importantly, these approaches are only effective when workplace leaders signal that risk mitigation takes equal precedence to commercial and client goals \cite{culbertson2025organizational, madaio2020co}, and integrate these practices throughout product development, as exemplified by privacy-by-design approaches \cite{cavoukian2009privacy}.

\subsection{Improving Ability to Address Agentic AI Risks}


Our RQ3 findings identified a capability vs. risk control tension in agentic AI development: mitigating agentic AI risks often requires constraining the same characteristics that make products agentic.
Participants felt ill-equipped to assess agentic AI risks and implement controls consistently, though this gap does not strictly stem from the absence of agentic AI risk resources. 
Emerging tools and frameworks (e.g., \cite{credo2026_ai_governance,bagehorn_ai_2025,madkour2026_agentic_profile}) support some aspects of agentic AI risk identification, governance, and mitigation.
Two RQ3 findings illustrate this gap: 1) developers lacked reliable ground truth for assessing agentic risks, and 2) many controls --- validator agents, LLM-as-a-judge --- were themselves agentic and subject to the same risks they were meant to mitigate. 
In short, developers need stronger support for navigating trade-offs between capability vs. risk control, as well as for deciding which controls are appropriate and feasible for their use case.

HAI research offers a useful precedent. Artifacts such as checklists (e.g., \cite{madaio2020co}), model cards (e.g., \cite{mitchell2019model}), and impact assessments (e.g., \cite{fiesler2018participant}) have helped practitioners translate high-level principles such as fairness, accountability, and transparency into actionable, practitioner-facing workflows for documenting, evaluating, and reflecting on AI systems. 
Future agentic AI design artifacts can build on this approach by helping teams articulate what agentic characteristics their products rely on, what risks those characteristics create, and what trade-offs different controls would introduce --- including whether the control is itself agentic, and thus vulnerable to the same failure modes as the system it governs.

\subsection{Limitations}
First, because our findings are grounded in participants' reported experiences, they should not be interpreted as representative of all agentic AI developers or industry contexts. 
We also did not explicitly prime participants on what should or should not count as agentic AI risks or practices. 
Participants may therefore have responded differently under a more heavily scaffolded protocol.
Second, because we recruited developers who already had experience addressing risks in user-facing agentic AI products, our sample may be skewed toward participants with relatively high risk awareness. 
Third, our sample was drawn from a single company, allowing us to study the emergent, contested phenomenon in depth, but also limiting generalizability. 

\section{Conclusion}
We interviewed 35 industry developers who build user-facing agentic AI products to examine how they perceive, prioritize, and address emergent risks in their products. 
Developers perceived agentic AI as creating and exacerbating risks above and beyond standard generative AI products. 
They perceived and prioritized risks through a product- and business-oriented model of success, focusing on risks tied to agent performance and business objectives, and de-emphasizing risks that introduced opportunity costs or fell outside of their perceived scope of work. 
Developers also implemented layered controls to constrain agent behavior, reduce agentic errors, strengthen access control, and support human-agent interaction, but these controls often worked by constraining the same characteristics that made their systems agentic. 
Together, our findings reveal a capability vs. risk control tension in agentic AI development: developers need to manage risks that emerge from agentic capabilities, but their current tools, incentives, and organizational structures make risk mitigation difficult to operationalize in practice. 

\section*{Acknowledgements}
We thank Justin Weisz, Mike Hind, and the entire Human-AI Collaboration team for their feedback on this project. We also thank Isadora Krsek and Sijia Xiao for their feedback on the manuscript. Finally, we thank the reviewers for their constructive feedback. Lee and Das's contributions to this project were supported, in part, by NSF SaTC grant \#2316768 and the CyLab Presidential Fellowship.

\bibliography{aaai2026,zotero}

@book{corbin2008basics,
  author    = {Corbin, Juliet and Strauss, Anselm},
  title     = {Basics of Qualitative Research: Techniques and Procedures for Developing Grounded Theory},
  edition   = {3},
  publisher = {SAGE Publications},
  address   = {Thousand Oaks, CA},
  year      = {2008},
  isbn      = {9781412906449}
}

@article{das2022security,
  title={The Security \& Privacy Acceptance Framework (SPAF)},
  author={Das, Sauvik and Faklaris, Cori and Hong, Jason I and Dabbish, Laura A and others},
  journal={Foundations and Trends{\textregistered} in Privacy and Security},
  volume={5},
  number={1-2},
  pages={1--143},
  year={2022},
  publisher={Now Publishers, Inc.}
}

@misc{roth2026openclawskills,
  author       = {Roth, Emma},
  title        = {{OpenClaw}'s {AI} ``Skill'' Extensions Are a Security Nightmare},
  year         = {2026},
  month        = feb,
  howpublished = {\url{https://www.theverge.com/news/874011/openclaw-ai-skill-clawhub-extensions-security-nightmare}},
  note         = {Published February 4, 2026. Accessed: 2026-04-25}
}

@inproceedings{kelley_there_2023,
  author    = {Patrick Gage Kelley and Celestina Cornejo and Lisa Hayes and Ellie Shuo Jin and Aaron Sedley and Kurt Thomas and Yongwei Yang and Allison Woodruff},
  title     = {{``There will be less privacy, of course'': How and why people in 10 countries expect {AI} will affect privacy in the future}},
  booktitle = {Nineteenth Symposium on Usable Privacy and Security ({SOUPS} 2023)},
  year      = {2023},
  pages     = {579--603},
  address   = {Anaheim, CA},
  publisher = {{USENIX} Association},
  isbn      = {978-1-939133-36-6},
  url       = {https://www.usenix.org/conference/soups2023/presentation/kelley},
  month     = aug
}

@inproceedings{tahaei2021privacy,
  title={Privacy champions in software teams: understanding their motivations, strategies, and challenges},
  author={Tahaei, Mohammad and Frik, Alisa and Vaniea, Kami},
  booktitle={Proceedings of the 2021 CHI Conference on Human Factors in Computing Systems},
  pages={1--15},
  year={2021}
}

@inproceedings {lee2024idontknown,
author = {Hao-Ping (Hank) Lee and Lan Gao and Stephanie Yang and Jodi Forlizzi and Sauvik Das},
title = {"I Don{\textquoteright}t Know If We{\textquoteright}re Doing Good. I Don{\textquoteright}t Know If We{\textquoteright}re Doing Bad": Investigating How Practitioners Scope, Motivate, and Conduct Privacy Work When Developing {AI} Products},
booktitle = {33rd USENIX Security Symposium (USENIX Security 24)},
year = {2024},
isbn = {978-1-939133-44-1},
address = {Philadelphia, PA},
pages = {4873--4890},
url = {https://www.usenix.org/conference/usenixsecurity24/presentation/lee},
publisher = {USENIX Association},
month = aug
}

@article{yao2022react,
  title={React: Synergizing reasoning and acting in language models},
  author={Yao, Shunyu and Zhao, Jeffrey and Yu, Dian and Du, Nan and Shafran, Izhak and Narasimhan, Karthik and Cao, Yuan},
  journal={arXiv preprint arXiv:2210.03629},
  year={2022}
}

@article{orr2020attributions,
  title={Attributions of ethical responsibility by Artificial Intelligence practitioners},
  author={Orr, Will and Davis, Jenny L.},
  journal={Information, Communication \& Society},
  volume={23},
  number={5},
  pages={719--735},
  year={2020},
  publisher={Taylor \& Francis},
  doi={10.1080/1369118X.2020.1713842}
}

@article{rakova2021where,
  title={Where Responsible AI meets Reality: Practitioner Perspectives on Enablers for Shifting Organizational Practices},
  author={Rakova, Bogdana and Yang, Jingying and Cramer, Henriette and Chowdhury, Rumman},
  journal={Proceedings of the ACM on Human-Computer Interaction},
  volume={5},
  number={CSCW1},
  articleno={7},
  pages={1--23},
  year={2021},
  doi={10.1145/3449081}
}

@article{cavoukian2009privacy,
  title={Privacy by design: The 7 foundational principles},
  author={Cavoukian, Ann and others},
  journal={Information and privacy commissioner of Ontario, Canada},
  volume={5},
  number={2009},
  pages={12},
  year={2009}
}

@article{shneiderman2020bridging,
  title={Bridging the gap between ethics and practice: guidelines for reliable, safe, and trustworthy human-centered AI systems},
  author={Shneiderman, Ben},
  journal={ACM Transactions on Interactive Intelligent Systems (TiiS)},
  volume={10},
  number={4},
  pages={1--31},
  year={2020},
  publisher={ACM New York, NY, USA}
}

@article{winfield2018ethical,
  title={Ethical governance is essential to building trust in robotics and artificial intelligence systems},
  author={Winfield, Alan FT and Jirotka, Marina},
  journal={Philosophical Transactions of the Royal Society A: Mathematical, Physical and Engineering Sciences},
  volume={376},
  number={2133},
  pages={20180085},
  year={2018},
  publisher={The Royal Society Publishing}
}

@techreport{imda2026_agentic_mgf,
  author      = {{Infocomm Media Development Authority}},
  title       = {Model AI Governance Framework for Agentic AI},
  institution = {Infocomm Media Development Authority},
  type        = {Model Governance Framework},
  year        = {2026},
  month       = jan,
  address     = {Singapore},
  url         = {https://www.imda.gov.sg/-/media/imda/files/about/emerging-tech-and-research/artificial-intelligence/mgf-for-agentic-ai.pdf},
  urldate     = {2026-05-12},
  note        = {Version 1.0}
}

@techreport{wef2025_ai_agents,
  author      = {{World Economic Forum} and {Capgemini}},
  title       = {AI Agents in Action: Foundations for Evaluation and Governance},
  institution = {World Economic Forum},
  type        = {White Paper},
  year        = {2025},
  month       = nov,
  address     = {Geneva, Switzerland},
  url         = {https://www.weforum.org/publications/ai-agents-in-action-foundations-for-evaluation-and-governance/},
  urldate     = {2026-05-12}
}

@inproceedings{wang_farsight_2024,
	address = {New York, NY, USA},
	series = {{CHI} '24},
	title = {Farsight: {Fostering} {Responsible} {AI} {Awareness} {During} {AI} {Application} {Prototyping}},
	isbn = {979-8-4007-0330-0},
	shorttitle = {Farsight},
	url = {https://dl.acm.org/doi/10.1145/3613904.3642335},
	doi = {10.1145/3613904.3642335},
	urldate = {2025-08-23},
	booktitle = {Proceedings of the 2024 {CHI} {Conference} on {Human} {Factors} in {Computing} {Systems}},
	publisher = {Association for Computing Machinery},
	author = {Wang, Zijie J. and Kulkarni, Chinmay and Wilcox, Lauren and Terry, Michael and Madaio, Michael},
	month = may,
	year = {2024},
	pages = {1--40},
}

@inproceedings{lee2026privy,
author = {Lee, Hao-Ping (Hank) and Yang, Yu-Ju and Bilik, Matthew and Krsek, Isadora and von Davier, Thomas Serban and Monteiro, Kyzyl and Lin, Jason and Agarwal, Shivani and Forlizzi, Jodi and Das, Sauvik},
title = {Privy: Envisioning and Mitigating Privacy Risks for Consumer-facing AI Product Concepts},
year = {2026},
isbn = {9798400722783},
publisher = {Association for Computing Machinery},
address = {New York, NY, USA},
url = {https://doi.org/10.1145/3772318.3791279},
doi = {10.1145/3772318.3791279},
booktitle = {Proceedings of the 2026 CHI Conference on Human Factors in Computing Systems},
articleno = {1435},
numpages = {30},
location = {
},
series = {CHI '26}
}

@article{buccinca2023aha,
  title={Aha!: Facilitating ai impact assessment by generating examples of harms},
  author={Bu{\c{c}}inca, Zana and Pham, Chau Minh and Jakesch, Maurice and Ribeiro, Marco Tulio and Olteanu, Alexandra and Amershi, Saleema},
  journal={arXiv preprint arXiv:2306.03280},
  year={2023}
}

@article{ganguli2022red,
  title={Red teaming language models to reduce harms: Methods, scaling behaviors, and lessons learned},
  author={Ganguli, Deep and Lovitt, Liane and Kernion, Jackson and Askell, Amanda and Bai, Yuntao and Kadavath, Saurav and Mann, Ben and Perez, Ethan and Schiefer, Nicholas and Ndousse, Kamal and others},
  journal={arXiv preprint arXiv:2209.07858},
  year={2022}
}

@inproceedings{nahar2026i,
author = {Nahar, Nadia and Yang, Chenyang and Chen, Yanxin and Hanwen Deng, Wesley and Holstein, Ken and Eslami, Motahhare and K\"{a}stner, Christian},
title = {“I Don’t Think RAI Applies to My Model” – Engaging Non-champions with Sticky Stories for Responsible AI Work},
year = {2026},
isbn = {9798400722783},
publisher = {Association for Computing Machinery},
address = {New York, NY, USA},
url = {https://doi.org/10.1145/3772318.3791285},
doi = {10.1145/3772318.3791285},
booktitle = {Proceedings of the 2026 CHI Conference on Human Factors in Computing Systems},
articleno = {1373},
numpages = {23},
location = {
},
series = {CHI '26}
}

@inproceedings{chen2024an,
author = {Chen, Chaoran and Li, Weijun and Song, Wenxin and Ye, Yanfang and Yao, Yaxing and Li, Toby Jia-Jun},
title = {An Empathy-Based Sandbox Approach to Bridge the Privacy Gap among Attitudes, Goals, Knowledge, and Behaviors},
year = {2024},
isbn = {9798400703300},
publisher = {Association for Computing Machinery},
address = {New York, NY, USA},
url = {https://doi.org/10.1145/3613904.3642363},
doi = {10.1145/3613904.3642363},
booktitle = {Proceedings of the 2024 CHI Conference on Human Factors in Computing Systems},
articleno = {234},
numpages = {28},
location = {Honolulu, HI, USA},
series = {CHI '24}
}

@article{gebru2021datasheets,
  title={Datasheets for datasets},
  author={Gebru, Timnit and Morgenstern, Jamie and Vecchione, Briana and Vaughan, Jennifer Wortman and Wallach, Hanna and Iii, Hal Daum{\'e} and Crawford, Kate},
  journal={Communications of the ACM},
  volume={64},
  number={12},
  pages={86--92},
  year={2021},
  publisher={ACM New York, NY, USA}
}

@inproceedings{madaio2020co,
  title={Co-designing checklists to understand organizational challenges and opportunities around fairness in AI},
  author={Madaio, Michael A and Stark, Luke and Wortman Vaughan, Jennifer and Wallach, Hanna},
  booktitle={Proceedings of the 2020 CHI Conference on Human Factors in Computing Systems},
  pages={1--14},
  year={2020}
}

@inproceedings{mitchell2019model,
  title={Model cards for model reporting},
  author={Mitchell, Margaret and Wu, Simone and Zaldivar, Andrew and Barnes, Parker and Vasserman, Lucy and Hutchinson, Ben and Spitzer, Elena and Raji, Inioluwa Deborah and Gebru, Timnit},
  booktitle={Proceedings of the conference on fairness, accountability, and transparency},
  pages={220--229},
  year={2019}
}

@article{fiesler2018participant,
  title={“Participant” perceptions of Twitter research ethics},
  author={Fiesler, Casey and Proferes, Nicholas},
  journal={Social Media+ Society},
  volume={4},
  number={1},
  pages={2056305118763366},
  year={2018},
  publisher={SAGE Publications Sage UK: London, England}
}

@inproceedings{ruan2024identifying,
  title={Identifying the Risks of LM Agents with an LM-Emulated Sandbox},
  author={Ruan, Yangjun and Maddison, Chris J. and Hashimoto, Tatsunori B. and Liang, Percy},
  booktitle={Proceedings of the Twelfth International Conference on Learning Representations (ICLR)},
  year={2024}
}

@inproceedings{debenedetti2024agentdojo,
  title={AgentDojo: A Dynamic Environment to Evaluate Prompt Injection Attacks and Defenses for LLM Agents},
  author={Debenedetti, Edoardo and Wang, Tianyu and Li, Jonathan and Feng, Zhaoyuan and Zhang, Yifan and Eger, Steffen and Tram{\`e}r, Florian},
  booktitle={Advances in Neural Information Processing Systems (NeurIPS)},
  year={2024}
}

@inproceedings{deng2024pentestgpt,
  title={PentestGPT: Evaluating and Harnessing Large Language Models for Automated Penetration Testing},
  author={Deng, Gelei and Liu, Yi and Mayoral-Vilches, Victor and Wang, Peng and Liu, Zhiyun Qian and Zhang, Yao and Liu, Yang and Song, Dawn},
  booktitle={33rd USENIX Security Symposium (USENIX Security 24)},
  pages={747--764},
  year={2024},
  publisher={USENIX Association}
}

@article{crispino2023agent,
  title={Agent instructs large language models to be general zero-shot reasoners},
  author={Crispino, Nicholas and Montgomery, Kyle and Zeng, Fankun and Song, Dawn and Wang, Chenguang},
  journal={arXiv preprint arXiv:2310.03710},
  year={2023}
}

@inproceedings{agashe2025agent,
  title={Agent s: An open agentic framework that uses computers like a human},
  author={Agashe, Saaket and Han, Jiuzhou and Gan, Shuyu and Yang, Jiachen and Li, Ang and Wang, Xin},
  booktitle={International Conference on Learning Representations},
  volume={2025},
  pages={22924--22946},
  year={2025}
}

@article{culbertson2025organizational,
  title   = {How Does Organizational Culture Influence the Adoption of AI Ethics Practices Across Sectors?},
  author  = {Culbertson, Shane},
  journal = {Muma Business Review},
  volume  = {9},
  number  = {18},
  pages   = {193--210},
  year    = {2025},
  month   = nov,
  url     = {https://mumabusinessreview.org/2025/MBR-09-18-193-210-Culbertson-EthicalAI.pdf}
}

@inproceedings{lee2024taxonomy,
author = {Lee, Hao-Ping (Hank) and Yang, Yu-Ju and Von Davier, Thomas Serban and Forlizzi, Jodi and Das, Sauvik},
title = {Deepfakes, Phrenology, Surveillance, and More! A Taxonomy of AI Privacy Risks},
year = {2024},
isbn = {9798400703300},
publisher = {Association for Computing Machinery},
address = {New York, NY, USA},
url = {https://doi.org/10.1145/3613904.3642116},
doi = {10.1145/3613904.3642116},
booktitle = {Proceedings of the 2024 CHI Conference on Human Factors in Computing Systems},
articleno = {775},
numpages = {19},
location = {Honolulu, HI, USA},
series = {CHI '24}
}

@article{Gan2026Navigating,
author = {Gan, Yuyou and Yang, Yong and Ma, Zhe and He, Ping and Zeng, Rui and Wang, Yiming and Li, Qingming and Zhou, Chunyi and Li, Songze and Wang, Ting and Gao, Yunjun and Wu, Yingcai and Ji, Shouling},
title = {Navigating the Risks: A Survey of Security and Privacy Threats in LLM-Based Agents},
year = {2026},
publisher = {Association for Computing Machinery},
address = {New York, NY, USA},
issn = {1049-331X},
url = {https://doi.org/10.1145/3807666},
doi = {10.1145/3807666},
note = {Just Accepted},
journal = {ACM Trans. Softw. Eng. Methodol.},
month = may
}

@inproceedings{Chang2026SystematicLit,
author = {Chang, Tyler and Razi, Afsaneh},
title = {A Systematic Literature Review of Generative AI Risks and Harms for Youth},
year = {2026},
isbn = {9798400722813},
publisher = {Association for Computing Machinery},
address = {New York, NY, USA},
url = {https://doi.org/10.1145/3772363.3798673},
doi = {10.1145/3772363.3798673},
booktitle = {Proceedings of the Extended Abstracts of the 2026 CHI Conference on Human Factors in Computing Systems},
articleno = {54},
numpages = {22},
location = {
},
series = {CHI EA '26}
}

@inproceedings{Steenstra2025RiskOntology,
author = {Steenstra, Ian and Bickmore, Timothy},
title = {A Risk Ontology for Evaluating AI-Powered Psychotherapy Virtual Agents},
year = {2025},
isbn = {9798400715082},
publisher = {Association for Computing Machinery},
address = {New York, NY, USA},
url = {https://doi.org/10.1145/3717511.3749286},
doi = {10.1145/3717511.3749286},
booktitle = {Proceedings of the 25th ACM International Conference on Intelligent Virtual Agents},
articleno = {32},
numpages = {4},
location = {
},
series = {IVA '25}
}

@inproceedings{li2026characterizingRisks,
author = {Li, Lingyao and Ma, Renkai and Xue, Zhaoqian and Xiong, Junjie},
title = {Characterizing User-Reported Risks across LLM Chatbots},
year = {2026},
isbn = {9798400722783},
publisher = {Association for Computing Machinery},
address = {New York, NY, USA},
url = {https://doi.org/10.1145/3772318.3791505},
doi = {10.1145/3772318.3791505},
booktitle = {Proceedings of the 2026 CHI Conference on Human Factors in Computing Systems},
articleno = {602},
numpages = {26},
location = {
},
series = {CHI '26}
}

@techreport{bellogin2025ACM_Europe,
 author       = {Bellogín, Alejandro and Giudici, Paolo and Larsson, Stefan and Pang, Jun and Schimpf, Gerhard and Sengupta, Biswa and Solmaz, Gürkan},
  institution  = {{Association for Computing Machinery (ACM)}},
  language     = {{eng}},
  month        = {{10}},
  series       = {{ACM Europe TPC - Autonomous Systems Subcommittee}},
  title        = {{Systemic Risks Associated with Agentic AI : A Policy Brief}},
  url          = {{https://lup.lub.lu.se/search/files/231065300/Bellog_n_et_al_2025_Systemic_Risks_Associated_with_Agentic_AI_A_Policy_Brief.pdf}},
  year         = {{2025}},
}

@Inbook{Gangavarapu2025,
author="Gangavarapu, Rajendra",
title="AI Governance: Preparing for the Rise of Agentic AI",
bookTitle="Mastering AI Governance: A Guide to Building Trustworthy and Transparent AI Systems",
year="2025",
publisher="Springer Nature Switzerland",
address="Cham",
pages="111--119",
isbn="978-3-031-93681-4",
doi="10.1007/978-3-031-93681-4_12",
url="https://doi.org/10.1007/978-3-031-93681-4_12"
}

@inproceedings{condon2026Bias_Eval,
author = {Condon, Gary and Jilani, Musfira},
title = {Towards a framework for Bias Evaluation of AI Agents in Agentic Workflows},
year = {2026},
isbn = {9798400721533},
publisher = {Association for Computing Machinery},
address = {New York, NY, USA},
url = {https://doi.org/10.1145/3777490.3777504},
doi = {10.1145/3777490.3777504},
booktitle = {Proceedings of the 2026 Conference on Human Centred Artificial Intelligence - Education and Practice},
pages = {46--52},
numpages = {7},
location = {
},
series = {HCAIep '26}
}

@article{glaser2024governance,
  title        = {Governance of artificial intelligence and machine learning in pharmacovigilance: what works today and what more is needed?},
  author       = {Glaser, M. and Littlebury, R.},
  journal      = {Therapeutic Advances in Drug Safety},
  volume       = {15},
  pages        = {20420986241293303},
  year         = {2024},
  doi          = {10.1177/20420986241293303},
  url          = {https://doi.org/10.1177/20420986241293303}
}

@inproceedings{Kenthapadi2023GenAI_and_RAI,
author = {Kenthapadi, Krishnaram and Lakkaraju, Himabindu and Rajani, Nazneen},
title = {Generative AI meets Responsible AI: Practical Challenges and Opportunities},
year = {2023},
isbn = {9798400701030},
publisher = {Association for Computing Machinery},
address = {New York, NY, USA},
url = {https://doi.org/10.1145/3580305.3599557},
doi = {10.1145/3580305.3599557},
booktitle = {Proceedings of the 29th ACM SIGKDD Conference on Knowledge Discovery and Data Mining},
pages = {5805--5806},
numpages = {2},
location = {Long Beach, CA, USA},
series = {KDD '23}
}

@article{Li2023TrustworthyAI,
author = {Li, Bo and Qi, Peng and Liu, Bo and Di, Shuai and Liu, Jingen and Pei, Jiquan and Yi, Jinfeng and Zhou, Bowen},
title = {Trustworthy AI: From Principles to Practices},
year = {2023},
issue_date = {September 2023},
publisher = {Association for Computing Machinery},
address = {New York, NY, USA},
volume = {55},
number = {9},
issn = {0360-0300},
url = {https://doi.org/10.1145/3555803},
doi = {10.1145/3555803},
journal = {ACM Comput. Surv.},
month = jan,
articleno = {177},
numpages = {46}
}

@article{thiebes2021trustworthy,
  title        = {Trustworthy Artificial Intelligence},
  author       = {Thiebes, S. and Lins, S. and Sunyaev, A.},
  journal      = {Electronic Markets},
  volume       = {31},
  number       = {2},
  pages        = {447--464},
  year         = {2021},
  doi          = {10.1007/s12525-020-00441-4},
  url          = {https://doi.org/10.1007/s12525-020-00441-4}
}

@article{floridi2018ai4people,
  title        = {AI4People---An Ethical Framework for a Good AI Society: Opportunities, Risks, Principles, and Recommendations},
  author       = {Floridi, Luciano and Cowls, Josh and Beltrametti, Monica and Chatila, Raja and Chazerand, Patrice and Dignum, Virginia and Luetge, Christoph and Madelin, Robert and Pagallo, Ugo and Rossi, Francesca and Schafer, Burkhard and Valcke, Peggy and Vayena, Effy},
  journal      = {Minds and Machines},
  volume       = {28},
  number       = {4},
  pages        = {689--707},
  year         = {2018},
  doi          = {10.1007/s11023-018-9482-5},
  url          = {https://doi.org/10.1007/s11023-018-9482-5}
}

@article{Li2018Coconut,
author = {Li, Tianshi and Agarwal, Yuvraj and Hong, Jason I.},
title = {Coconut: An IDE Plugin for Developing Privacy-Friendly Apps},
year = {2018},
issue_date = {December 2018},
publisher = {Association for Computing Machinery},
address = {New York, NY, USA},
volume = {2},
number = {4},
url = {https://doi.org/10.1145/3287056},
doi = {10.1145/3287056},
journal = {Proc. ACM Interact. Mob. Wearable Ubiquitous Technol.},
month = dec,
articleno = {178},
numpages = {35}
}

@inproceedings{Li2022Developer_challenges,
author = {Li, Tianshi and Reiman, Kayla and Agarwal, Yuvraj and Cranor, Lorrie Faith and Hong, Jason I.},
title = {Understanding Challenges for Developers to Create Accurate Privacy Nutrition Labels},
year = {2022},
isbn = {9781450391573},
publisher = {Association for Computing Machinery},
address = {New York, NY, USA},
url = {https://doi.org/10.1145/3491102.3502012},
doi = {10.1145/3491102.3502012},
booktitle = {Proceedings of the 2022 CHI Conference on Human Factors in Computing Systems},
articleno = {588},
numpages = {24},
location = {New Orleans, LA, USA},
series = {CHI '22}
}

@inproceedings{ehsan2026HCXAI,
author = {Ehsan, Upol and Alabdulkarim, Amal and Holstein, Kenneth and Lee, Min Kyung and Riener, Andreas and Weisz, Justin D.},
title = {Human-Centered Explainable AI (HCXAI): Re-examining XAI in the Era of Agentic AI},
year = {2026},
isbn = {9798400722813},
publisher = {Association for Computing Machinery},
address = {New York, NY, USA},
url = {https://doi.org/10.1145/3772363.3778728},
doi = {10.1145/3772363.3778728},
booktitle = {Proceedings of the Extended Abstracts of the 2026 CHI Conference on Human Factors in Computing Systems},
articleno = {955},
numpages = {6},
location = {
},
series = {CHI EA '26}
}

@techreport{madkour2026_agentic_profile,
  author      = {Nada Madkour and Jessica Newman and Deepika Raman and Krystal Jackson and Evan R. Murphy and Charlotte Yuan},
  title       = {Agentic AI Risk-Management Standards Profile},
  institution = {UC Berkeley Center for Long-Term Cybersecurity},
  year        = {2026},
  month       = feb,
  address     = {Berkeley, CA},
  url         = {https://cltc.berkeley.edu/wp-content/uploads/2026/02/Agentic-AI-Risk-Management-Standards-Profile.pdf},
  note        = {Accessed: 2026-05-19}
}

@misc{credo2026_ai_governance,
  author       = {{Credo AI}},
  title        = {Credo AI: The Trusted Leader in AI Governance},
  year         = {2026},
  howpublished = {\url{https://www.credo.ai/}},
  note         = {Accessed: 2026-05-19}
}

@inproceedings{winston2025taxonomy,
  title={A taxonomy of failures in tool-augmented llms},
  author={Winston, Cailin and Just, Ren{\'e}},
  booktitle={2025 IEEE/ACM International Conference on Automation of Software Test (AST)},
  pages={125--135},
  year={2025},
  organization={IEEE}
}

@misc{hart2026kiro,
  author       = {Hart, Robert},
  title        = {Amazon Blames Human Employees for an {AI} Coding Agent's Mistake},
  howpublished = {\emph{The Verge}},
  year         = {2026},
  month        = feb,
  day          = {20},
  url          = {https://www.theverge.com/ai-artificial-intelligence/882005/amazon-blames-human-employees-for-an-ai-coding-agents-mistake},
  note         = {Published February 20, 2026. Accessed: May 19, 2026}
}

@misc{hart2026lobster,
  author       = {Hart, Robert},
  title        = {The {AI} Security Nightmare Is Here and It Looks Suspiciously Like Lobster},
  howpublished = {\emph{The Verge}},
  year         = {2026},
  month        = feb,
  day          = {19},
  url          = {https://www.theverge.com/ai-artificial-intelligence/881574/cline-openclaw-prompt-injection-hack},
  note         = {Published February 19, 2026. Accessed: April 25, 2026}
}

@misc{miehling_agentic_2025,
	title = {Agentic {AI} {Needs} a {Systems} {Theory}},
	url = {http://arxiv.org/abs/2503.00237},
	doi = {10.48550/arXiv.2503.00237},
	urldate = {2026-05-16},
	publisher = {arXiv},
	author = {Miehling, Erik and Ramamurthy, Karthikeyan Natesan and Varshney, Kush R. and Riemer, Matthew and Bouneffouf, Djallel and Richards, John T. and Dhurandhar, Amit and Daly, Elizabeth M. and Hind, Michael and Sattigeri, Prasanna and Wei, Dennis and Rawat, Ambrish and Gajcin, Jasmina and Geyer, Werner},
	month = feb,
	year = {2025},
	note = {arXiv:2503.00237 [cs.AI]},
}

@inproceedings{shome_why_2026,
	title = {Why {Johnny} {Can}’t {Use} {Agents}: {Industry} {Aspirations} vs. {User} {Realities} with {AI} {Agents}},
	language = {en},
	booktitle = {{ACM} {Conference} on {AI} and {Agentic} {Systems} ({CAIS})},
	author = {Shome, Pradyumna and Krishnan, Sashreek and Das, Sauvik},
	year = {2026},
}

@article{acharya_agentic_2025,
	title = {Agentic {AI}: {Autonomous} {Intelligence} for {Complex} {Goals}—{A} {Comprehensive} {Survey}},
	volume = {13},
	issn = {2169-3536},
	shorttitle = {Agentic {AI}},
	url = {https://ieeexplore.ieee.org/abstract/document/10849561},
	doi = {10.1109/ACCESS.2025.3532853},
	urldate = {2026-05-16},
	journal = {IEEE Access},
	author = {Acharya, Deepak Bhaskar and Kuppan, Karthigeyan and Divya, B.},
	year = {2025},
	pages = {18912--18936},
}

@article{wang_survey_2024,
	title = {A survey on large language model based autonomous agents},
	volume = {18},
	issn = {2095-2236},
	url = {https://doi.org/10.1007/s11704-024-40231-1},
	doi = {10.1007/s11704-024-40231-1},
	language = {en},
	number = {6},
	urldate = {2026-05-12},
	journal = {Frontiers of Computer Science},
	author = {Wang, Lei and Ma, Chen and Feng, Xueyang and Zhang, Zeyu and Yang, Hao and Zhang, Jingsen and Chen, Zhiyuan and Tang, Jiakai and Chen, Xu and Lin, Yankai and Zhao, Wayne Xin and Wei, Zhewei and Wen, Jirong},
	month = mar,
	year = {2024},
	pages = {186345},
}

@misc{pan_measuring_2026,
	title = {Measuring {Agents} in {Production}},
	url = {http://arxiv.org/abs/2512.04123},
	doi = {10.48550/arXiv.2512.04123},
	urldate = {2026-05-01},
	publisher = {arXiv},
	author = {Pan, Melissa Z. and Arabzadeh, Negar and Cogo, Riccardo and Zhu, Yuxuan and Xiong, Alexander and Agrawal, Lakshya A. and Mao, Huanzhi and Shen, Emma and Pallerla, Sid and Patel, Liana and Liu, Shu and Shi, Tianneng and Liu, Xiaoyuan and Davis, Jared Quincy and Lacavalla, Emmanuele and Basile, Alessandro and Yang, Shuyi and Castro, Paul and Kang, Daniel and Gonzalez, Joseph E. and Sen, Koushik and Song, Dawn and Stoica, Ion and Zaharia, Matei and Ellis, Marquita},
	month = feb,
	year = {2026},
	note = {arXiv:2512.04123 [cs]},
}

@misc{bagehorn_ai_2025,
	title = {{AI} {Risk} {Atlas}: {Taxonomy} and {Tooling} for {Navigating} {AI} {Risks} and {Resources}},
	shorttitle = {{AI} {Risk} {Atlas}},
	url = {https://arxiv.org/abs/2503.05780v2},
	language = {en},
	urldate = {2026-04-20},
	journal = {arXiv.org},
	author = {Bagehorn, Frank and Brimijoin, Kristina and Daly, Elizabeth M. and He, Jessica and Hind, Michael and Garces-Erice, Luis and Giblin, Christopher and Giurgiu, Ioana and Martino, Jacquelyn and Nair, Rahul and Piorkowski, David and Rawat, Ambrish and Richards, John and Rooney, Sean and Salwala, Dhaval and Tirupathi, Seshu and Urbanetz, Peter and Varshney, Kush R. and Vejsbjerg, Inge and Wolf-Bauwens, Mira L.},
	month = feb,
	year = {2025},
}

@article{sanderson_ai_2023,
	title = {{AI} {Ethics} {Principles} in {Practice}: {Perspectives} of {Designers} and {Developers}},
	volume = {4},
	issn = {2637-6415},
	shorttitle = {{AI} {Ethics} {Principles} in {Practice}},
	url = {http://arxiv.org/abs/2112.07467},
	doi = {10.1109/TTS.2023.3257303},
	number = {2},
	urldate = {2025-05-28},
	journal = {IEEE Transactions on Technology and Society},
	author = {Sanderson, Conrad and Douglas, David and Lu, Qinghua and Schleiger, Emma and Whittle, Jon and Lacey, Justine and Newnham, Glenn and Hajkowicz, Stefan and Robinson, Cathy and Hansen, David},
	month = jun,
	year = {2023},
	note = {arXiv:2112.07467 [cs]},
	pages = {171--187},
}

@misc{koessler_risk_2023,
	title = {Risk assessment at {AGI} companies: {A} review of popular risk assessment techniques from other safety-critical industries},
	shorttitle = {Risk assessment at {AGI} companies},
	url = {http://arxiv.org/abs/2307.08823},
	doi = {10.48550/arXiv.2307.08823},
	urldate = {2025-05-22},
	publisher = {arXiv},
	author = {Koessler, Leonie and Schuett, Jonas},
	month = jul,
	year = {2023},
	note = {arXiv:2307.08823 [cs]},
}

@article{pant_ethics_2024,
	title = {Ethics in the {Age} of {AI}: {An} {Analysis} of {AI} {Practitioners}’ {Awareness} and {Challenges}},
	volume = {33},
	issn = {1049-331X},
	shorttitle = {Ethics in the {Age} of {AI}},
	url = {https://dl.acm.org/doi/10.1145/3635715},
	doi = {10.1145/3635715},
	number = {3},
	urldate = {2025-05-22},
	journal = {ACM Trans. Softw. Eng. Methodol.},
	author = {Pant, Aastha and Hoda, Rashina and Spiegler, Simone V. and Tantithamthavorn, Chakkrit and Turhan, Burak},
	month = mar,
	year = {2024},
	pages = {80:1--80:35},
}

@inproceedings{chan_harms_2023,
	address = {Chicago IL USA},
	title = {Harms from {Increasingly} {Agentic} {Algorithmic} {Systems}},
	isbn = {979-8-4007-0192-4},
	url = {https://dl.acm.org/doi/10.1145/3593013.3594033},
	doi = {10.1145/3593013.3594033},
	language = {en},
	urldate = {2025-05-22},
	booktitle = {2023 {ACM} {Conference} on {Fairness} {Accountability} and {Transparency}},
	publisher = {ACM},
	author = {Chan, Alan and Salganik, Rebecca and Markelius, Alva and Pang, Chris and Rajkumar, Nitarshan and Krasheninnikov, Dmitrii and Langosco, Lauro and He, Zhonghao and Duan, Yawen and Carroll, Micah and Lin, Michelle and Mayhew, Alex and Collins, Katherine and Molamohammadi, Maryam and Burden, John and Zhao, Wanru and Rismani, Shalaleh and Voudouris, Konstantinos and Bhatt, Umang and Weller, Adrian and Krueger, David and Maharaj, Tegan},
	month = jun,
	year = {2023},
	pages = {651--666},
}

@misc{kasirzadeh_characterizing_2025,
	title = {Characterizing {AI} {Agents} for {Alignment} and {Governance}},
	url = {http://arxiv.org/abs/2504.21848},
	doi = {10.48550/arXiv.2504.21848},
	urldate = {2025-05-21},
	publisher = {arXiv},
	author = {Kasirzadeh, Atoosa and Gabriel, Iason},
	month = apr,
	year = {2025},
	note = {arXiv:2504.21848 [cs]},
}

\appendix
\clearpage

\section{Appendix}
\subsection{Semi-structured Interview Protocol}

\subsubsection{Agentic AI Product Information}

\begin{enumerate}
    \item[1.] Can you briefly describe the AI agent you're working on?
    \item[2.] What is your role in the team?
    \item[3.] Who are the users of your AI agent?
\end{enumerate}

\subsubsection{Developers' Concerns for Their Agentic AI Products}

We'd like to learn more about your experience handling risks for your AI agent.

\begin{enumerate}
    \item[4.] (For each of the three HAI principles the participant selected in the pre-study questionnaire)
    \begin{enumerate}
        \item Can you recall the last time you and your team discussed a specific risk related to [the HAI principle]?
        \begin{enumerate}
            \item Why is [the risk mentioned] important to consider for your AI agent?
            \item Do you think [the risk mentioned] emerges because this product is agentic?
        \end{enumerate}
    \end{enumerate}
\end{enumerate}

\subsubsection{Agentic AI Risk Prioritization}

Here are the risks that you said are relevant to your AI agent. 
Now, spend a minute refreshing your memory with these risks, and rank these risks in the order of their importance for this particular AI agent.

(Wait for the participant to finish the risk-ranking activity.)

Now, I'd like you to reflect on how you ranked these risks.

\begin{enumerate}
    \item[5.] Let's start with the risk at the top, [the first risk]. Why is this risk ranked over all the other risks?
    \begin{enumerate}
        \item What did you consider when ranking this risk first?
    \end{enumerate}
    \item[6.] Let's take a look at the risk at the bottom, [the last risk]. Why is this risk ranked lower than all the other risks here?
    \begin{enumerate}
        \item What did you consider when ranking this risk last?
    \end{enumerate}
    \item[7.] Beyond [a summary of the factors the participant mentioned], is there anything else you considered when ranking these risks?
\end{enumerate}

\subsubsection{Bow-tie Analysis}

\paragraph{Introduction to the bow-tie analysis diagram}
Look at this bow-tie shape workspace: in the center, we will put the risk that we will be analyzing. 
On the left side, we look at what could cause the risk; these are called ``causes.'' 
On the right side, we think about what could happen after the risk happens; these are ``consequences.'' 
Then, we identify controls, things we can do to either prevent the risk from happening, or reduce and recover its impact if it does. 
This method helps us think about the underlying factors of risks and what we can do about them.

(The participant picks one of the ranked agentic AI risks that they have attempted to mitigate, and puts the risk at the center of the bow-tie diagram.)

\paragraph{Risk causes}

Now, let's first identify the root causes of the risk. 
These causes can be technical or non-technical.

\begin{enumerate}
    \item[8.] (Wait for the participant to put causes on the diagram, and for each cause, ask:)
    \begin{enumerate}
        \item How does this cause link to your AI agent?
    \end{enumerate}
\end{enumerate}

\paragraph{Preventive controls}

\begin{enumerate}
    \item[9.] (For each cause identified, ask:)
    \begin{enumerate}
        \item What preventive controls have you applied, either technical or non-technical, that can prevent [the cause] from causing [the risk]?
        \begin{enumerate}
            \item What challenges, if any, did you and your team face when applying this preventive control?
            \item What tradeoffs, if any, did you and your team have to make when applying this preventive control?
        \end{enumerate}
    \end{enumerate}
\end{enumerate}

\paragraph{Risk consequences}
Let's move on and think about potential consequences of this risk that you'd want to mitigate or reduce for your AI agent.

\begin{enumerate}
    \item[10.] (Wait for the participant to put consequences on the diagram, and for each consequence, ask:)
    \begin{enumerate}
        \item Who would be impacted?
        \item Why is the consequence crucial to consider?
    \end{enumerate}
\end{enumerate}

\paragraph{Protective controls}

\begin{enumerate}
    \item[11.] (For each consequence identified, ask:)
    \begin{enumerate}
        \item What protective controls have you applied, either technical or non-technical, that can reduce or prevent [the consequence] from happening?
        \begin{enumerate}
            \item What challenges, if any, did you and your team face when applying this protective control?
            \item What tradeoffs, if any, did you and your team have to make when applying this protective control?
        \end{enumerate}
    \end{enumerate}
\end{enumerate}

\subsubsection{Closing}

\begin{enumerate}
    \item[12.] Anything else you'd like to share about addressing risks for your AI agent before we wrap this up?
\end{enumerate}

\begin{figure*}
\centering
\includegraphics[width=1.5\columnwidth]{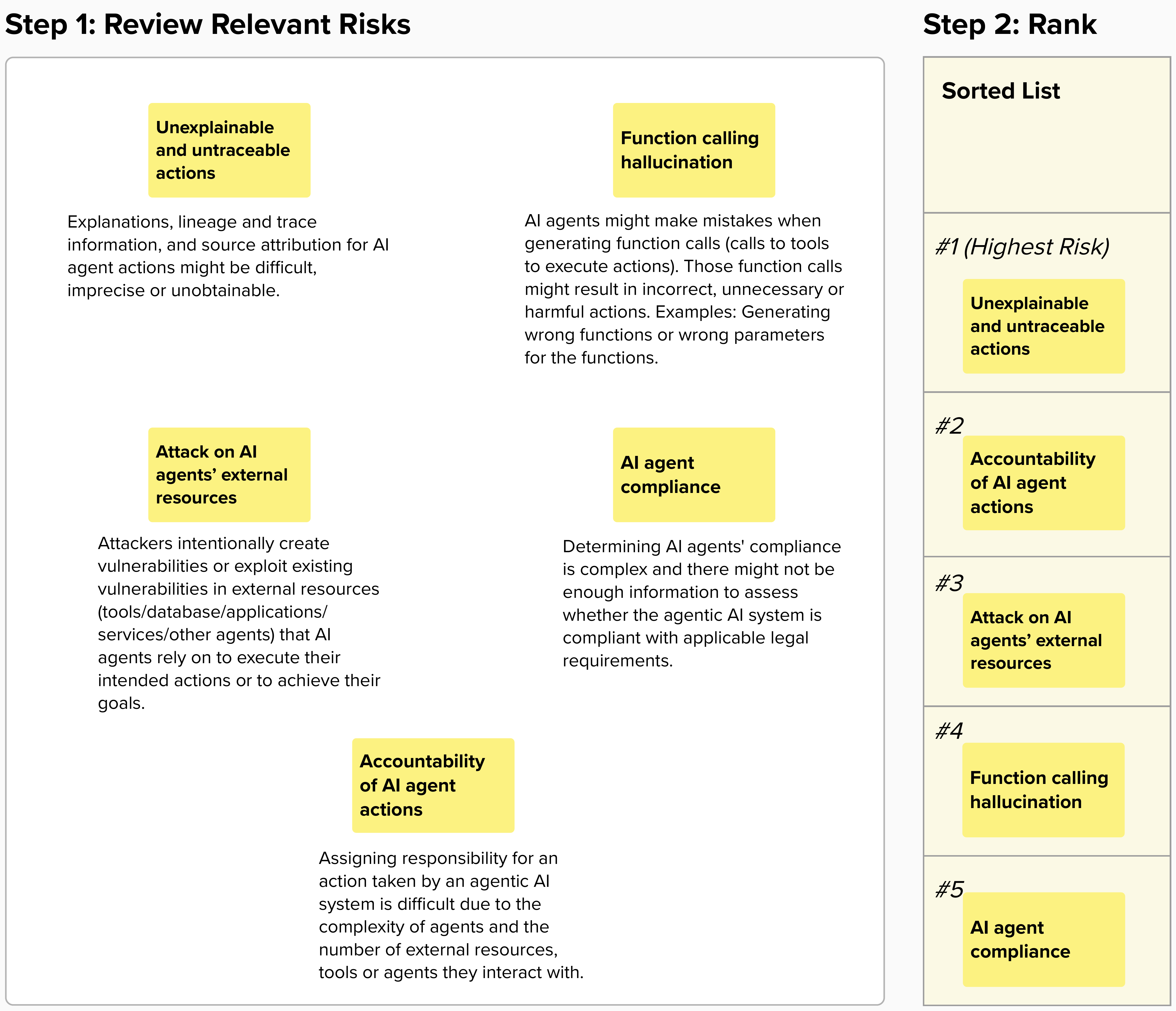}
\caption{Example of a completed risk-ranking activity used in the interview. Participants reviewed five agentic AI risks selected from their questionnaire responses, ranked them by how important they were to address for their agentic AI product, and explained the factors that shaped their ranking.}
\label{fig:risk ranking}
\end{figure*}

\begin{table*}
\centering
\caption{Agentic AI risks, grouped by human-centered AI (HAI) principle.}
\label{tab:appendix-agentic-ai-risk-table}
\arrayrulecolor[rgb]{0.8,0.8,0.8}
\resizebox{\linewidth}{!}{%
\begin{tabular}{|l|l|} 
\toprule
\textbf{HAI Principle} & \textbf{Agentic AI Risk \cite{bagehorn_ai_2025}} \\ 
\hline
\multirow{2}{*}{\begin{tabular}[c]{@{}l@{}}Privacy Protection\\and Security\end{tabular}} & \begin{tabular}[c]{@{}l@{}}\textbf{Sharing IP/PI/confidential information with user:} AI agents with unrestricted access to resources or databases or tools could potentially store\\and share PI/IP/confidential information with system users when performing their actions.\end{tabular} \\ 
\cline{2-2}
 & \begin{tabular}[c]{@{}l@{}}\textbf{Sharing IP/PI/confidential information with tools:} AI agents with unrestricted access to resources or databases or tools could potentially store and \\share PI/IP/confidential information with other tools or agents when performing their actions.\end{tabular} \\ 
\hline
\multirow{4}{*}{\begin{tabular}[c]{@{}l@{}}Reliability and\\Safety\end{tabular}} & \begin{tabular}[c]{@{}l@{}}\textbf{Attack on AI agents’ external resources:} Attackers intentionally create vulnerabilities or exploit existing vulnerabilities in external resources \\(tools/database/applications/services/other agents) that AI agents rely on to execute their intended actions or to achieve their goals.\end{tabular} \\ 
\cline{2-2}
 & \begin{tabular}[c]{@{}l@{}}\textbf{Unauthorized use:} If attackers can gain access to the AI agent and its components, they can perform actions that can have different levels of \\harm depending on the agent’s capabilities and information it has access to. Examples:\\- Using stored personal information to mimic identity or impersonate with an intent to deceive.\\- Manipulating AI agent’s behavior via feedback to the AI agent or corrupting its memory to change its behavior.\\- Manipulating the problem description or the goal to get the AI agent to behave badly or run harmful commands.\end{tabular} \\ 
\cline{2-2}
 & \begin{tabular}[c]{@{}l@{}}\textbf{Exploit trust mismatch:} Attackers might initiate injection attacks to bypass the trust boundary, which is a distinct point or conceptual line where \\the level of trust in a system, application or network changes. Background execution in multi-agent environments increases the risk of covert \\channels if input/output validation is weak.\end{tabular} \\ 
\cline{2-2}
 & \begin{tabular}[c]{@{}l@{}}\textbf{Function calling hallucination:} AI agents might make mistakes when generating function calls (calls to tools to execute actions). Those function \\calls might result in incorrect, unnecessary or harmful actions. Examples: Generating wrong functions or wrong parameters for the functions.\end{tabular} \\ 
\hline
\multirow{2}{*}{\begin{tabular}[c]{@{}l@{}}Transparency and\\Explainability\end{tabular}} & \begin{tabular}[c]{@{}l@{}}\textbf{Lack of AI agent transparency:} Lack of AI agent transparency is due to insufficient documentation of the AI agent design, development, \\evaluation process, absence of insights into the inner workings of the AI agent, and interaction with other agents/tools/resources.\end{tabular} \\ 
\cline{2-2}
 & \begin{tabular}[c]{@{}l@{}}\textbf{Unexplainable and untraceable actions:} Explanations, lineage and trace information, and source attribution for AI agent actions might be \\difficult, imprecise or unobtainable.\end{tabular} \\ 
\hline
\multirow{2}{*}{Fairness} & \begin{tabular}[c]{@{}l@{}}\textbf{Introduce data bias:} Specific actions taken by the AI agent, such as modifying a dataset or a database, can introduce bias in the resource that \\gets used by others or by itself to take actions.\end{tabular} \\ 
\cline{2-2}
 & \begin{tabular}[c]{@{}l@{}}\textbf{Discriminatory actions:} AI agents can take actions where one group of humans is unfairly advantaged over another due to the decisions of \\the model. This may be caused by AI agents’ biased actions that impact the world, in the resources consulted, and in the resource selection \\process. For example, an AI agent can generate code that can be biased.\end{tabular} \\ 
\hline
\multirow{4}{*}{\begin{tabular}[c]{@{}l@{}}Performance\\and Testing\end{tabular}} & \textbf{Incomplete AI agent evaluation:} Evaluating the performance or accuracy or an agent is difficult because of system complexity and open-endedness. \\ 
\cline{2-2}
 & \begin{tabular}[c]{@{}l@{}}\textbf{Redundant actions:} AI agents can execute actions that are not needed for achieving the goal.\\In an extreme case, AI agents might enter a cycle of executing the same actions repeatedly without any progress. This could happen because of \\unexpected conditions in the environment, the AI agent’s failure to reflect on its action, AI agent reasoning and planning errors or the AI agent’s \\lack of knowledge about the problem.\end{tabular} \\ 
\cline{2-2}
 & \begin{tabular}[c]{@{}l@{}}\textbf{Reproducibility:} Replicating agent behavior or output can be impacted by changes or updates made to external services and tools. This impact is \\increased if the agent is built with generative AI.\end{tabular} \\ 
\cline{2-2}
 & \begin{tabular}[c]{@{}l@{}}\textbf{Mitigation and maintenance:} The large number of components and dependencies that agent systems have complicates keeping them up to date \\and correcting problems.\end{tabular} \\ 
\hline
\multirow{2}{*}{Accountability} & \begin{tabular}[c]{@{}l@{}}\textbf{Accountability of AI agent actions:} Assigning responsibility for an action taken by an agentic AI system is difficult due to the complexity of agents \\and the number of external resources, tools or agents they interact with.\end{tabular} \\ 
\cline{2-2}
 & \begin{tabular}[c]{@{}l@{}}\textbf{AI agent compliance:} Determining AI agents' compliance is complex and there might not be enough information to assess whether the agentic AI \\system is compliant with applicable legal requirements.\end{tabular} \\ 
\hline
\multirow{2}{*}{\begin{tabular}[c]{@{}l@{}}Human-centered\\Values\end{tabular}} & \begin{tabular}[c]{@{}l@{}}\textbf{Over- or under-reliance on AI agents:} Reliance, that is the willingness to accept an AI agent behavior, depends on how much a user trusts that \\agent and what they are using it for. Over-reliance occurs when a user puts too much trust in an AI agent, accepting an AI agent’s behavior even \\when it is likely undesired. Under-reliance is the opposite, where the user doesn’t trust the AI agent but should.\\Increasing autonomy (to take action, select and consult resources/tools) of AI agents and the possibility of opaqueness and open-endedness increase\\the variability and visibility of agent behavior leading to difficulty in calibrating trust and possibly contributing to both over- and under-reliance.\end{tabular} \\ 
\cline{2-2}
 & \begin{tabular}[c]{@{}l@{}}\textbf{Misaligned actions:} AI agents can take actions that are not aligned with relevant human values, ethical considerations, guidelines and policies.\\Misaligned actions can occur in different ways such as:\\- Applying learned goals inappropriately to new or unforeseen situations.\\- Using AI agents for a purpose/goals that are beyond their intended use.\\- Selecting resources or tools in a biased way\\- Using deceptive tactics to achieve the goal by developing the capacity for scheming based on the instructions given within a specific context.\\- Compromising on AI agent values to work with another AI agent or tool to accomplish the task.\end{tabular} \\ 
\hline
\multirow{4}{*}{\begin{tabular}[c]{@{}l@{}}Human, Social and\\Environmental\\Wellbeing\end{tabular}} & \begin{tabular}[c]{@{}l@{}}\textbf{AI agents' impact on human dignity:} If human workers perceive AI agents as being better at doing the job of the human, the human can \\experience a decline in their self-worth and wellbeing.\end{tabular} \\ 
\cline{2-2}
 & \begin{tabular}[c]{@{}l@{}}\textbf{AI agents' impact on human agency:} The autonomous nature of AI agents in performing tasks or taking actions might affect the individuals’\\ability to engage in critical thinking, make choices and act independently.\end{tabular} \\ 
\cline{2-2}
 & \begin{tabular}[c]{@{}l@{}}\textbf{AI agents' impact on jobs:} Widespread adoption of AI agents to perform complex tasks might lead to widespread automation of roles and \\might lead to job displacement.\end{tabular} \\ 
\cline{2-2}
 & \begin{tabular}[c]{@{}l@{}}\textbf{AI agents' impact on environment:} Complexity of the tasks and possibility of AI agents performing redundant actions could lead to \\computational inefficiencies and add to the environmental impact.\end{tabular} \\
\bottomrule
\end{tabular}
}
\arrayrulecolor{black}
\end{table*}

\begin{table*}
\centering
\caption{Participant Information}
\label{tab:participant}
\arrayrulecolor[rgb]{0.8,0.8,0.8}
\resizebox{0.8\linewidth}{!}{%
\begin{tabular}{|l|l|l|l|l|} 
\toprule
\textbf{PID} & \textbf{Job title} & \begin{tabular}[c]{@{}l@{}}\textbf{Application domain }\\\textbf{of the agentic AI products}\end{tabular} & \textbf{Selected human-centered AI principles} & \textbf{Ranked agentic AI risks} \\ 
\hline
P1 & Technical Specialist & Sales \& Marketing & \begin{tabular}[c]{@{}l@{}}Reliability and Safety\\Privacy Protection and Security\\Performance and Testing\end{tabular} & \begin{tabular}[c]{@{}l@{}}Redundant actions\\Function calling hallucination\\Incomplete AI agent evaluation\\Attack on AI agents’ external resources\\Unauthorized use\end{tabular} \\ 
\hline
P2 & Research Scientist & Information Technology & \begin{tabular}[c]{@{}l@{}}Performance and Testing\\Reliability and Safety\\Transparency and Explainability\end{tabular} & \begin{tabular}[c]{@{}l@{}}Incomplete AI agent evaluation\\Function calling hallucination\\Redundant actions\\Mitigation and maintenance\\Exploit trust mismatch\end{tabular} \\ 
\hline
P3 & Software Developer & Research \& Development & \begin{tabular}[c]{@{}l@{}}Privacy Protection and Security\\Reliability and Safety\\Performance and Testing\end{tabular} & \begin{tabular}[c]{@{}l@{}}Sharing IP/PI/confidential information with tools\\Exploit trust mismatch\\Attack on AI agents’ external resources\\Function calling hallucination\\Unauthorized use\end{tabular} \\ 
\hline
P4 & Research Scientist & Information Technology & \begin{tabular}[c]{@{}l@{}}Transparency and Explainability\\Performance and Testing\\Accountability\end{tabular} & \begin{tabular}[c]{@{}l@{}}Incomplete AI agent evaluation\\Accountability of AI agent actions\\Unexplainable and untraceable actions\\Lack of AI agent transparency\\Redundant actions\end{tabular} \\ 
\hline
P5 & AI Engineer & Corporate Services & \begin{tabular}[c]{@{}l@{}}Reliability and Safety\\Transparency and Explainability\\Accountability\end{tabular} & \begin{tabular}[c]{@{}l@{}}Attack on AI agents’ external resources\\Unauthorized use\\Accountability of AI agent actions\\Lack of AI agent transparency\\AI agent compliance\end{tabular} \\ 
\hline
P6 & Software Developer & Software Development & \begin{tabular}[c]{@{}l@{}}Privacy Protection and Security\\Accountability\\Human-centred Values\end{tabular} & \begin{tabular}[c]{@{}l@{}}Sharing IP/PI/confidential information with user\\Misaligned actions\\AI agent compliance\\Sharing IP/PI/confidential information with tools\\Over- or under-reliance on AI agents\end{tabular} \\ 
\hline
P7 & Software Developer & Information Technology & \begin{tabular}[c]{@{}l@{}}Performance and Testing\\Reliability and Safety\\Transparency and Explainability\end{tabular} & \begin{tabular}[c]{@{}l@{}}Attack on AI agents’ external resources\\Unexplainable and untraceable actions\\Unauthorized use\\Redundant actions\\Incomplete AI agent evaluation\end{tabular} \\ 
\hline
P8 & Software Developer & Customer Support & \begin{tabular}[c]{@{}l@{}}Performance and Testing\\Transparency and Explainability\\Accountability\end{tabular} & \begin{tabular}[c]{@{}l@{}}Unexplainable and untraceable actions\\AI agent compliance\\Accountability of AI agent actions\\Incomplete AI agent evaluation\\Mitigation and maintenance\end{tabular} \\ 
\hline
P9 & Software Developer & Information Technology & \begin{tabular}[c]{@{}l@{}}Performance and Testing\\Privacy Protection and Security\\Reliability and Safety\end{tabular} & \begin{tabular}[c]{@{}l@{}}Incomplete AI agent evaluation\\Reproducibility\\Sharing IP/PI/confidential information with tools\\Function calling hallucination\\Redundant actions\end{tabular} \\ 
\hline
P10 & AI Engineer & Finance \& Banking & \begin{tabular}[c]{@{}l@{}}Reliability and Safety\\Transparency and Explainability\\Performance and Testing\end{tabular} & \begin{tabular}[c]{@{}l@{}}Reproducibility\\Lack of AI agent transparency\\Unexplainable and untraceable actions\\Function calling hallucination\\Attack on AI agents’ external resources\end{tabular} \\ 
\hline
P11 & Software Developer & Corporate Services & \begin{tabular}[c]{@{}l@{}}Privacy Protection and Security\\Fairness\\Transparency and Explainability\end{tabular} & \begin{tabular}[c]{@{}l@{}}Sharing IP/PI/confidential information with tools\\Unexplainable and untraceable actions\\Lack of AI agent transparency\\Introduce data bias\\Discriminatory actions\end{tabular} \\ 
\hline
P12 & Platform Engineer & Customer Support & \begin{tabular}[c]{@{}l@{}}Transparency and Explainability\\Performance and Testing\\Reliability and Safety\end{tabular} & \begin{tabular}[c]{@{}l@{}}Incomplete AI agent evaluation\\Reproducibility\\Unexplainable and untraceable actions\\Lack of AI agent transparency\\Function calling hallucination\end{tabular} \\ 
\hline
P13 & Software Developer & Corporate Services & \begin{tabular}[c]{@{}l@{}}Accountability\\Privacy Protection and Security\\Transparency and Explainability\end{tabular} & \begin{tabular}[c]{@{}l@{}}Sharing IP/PI/confidential information with user\\Accountability of AI agent actions\\Sharing IP/PI/confidential information with tools\\Lack of AI agent transparency\\AI agent compliance\end{tabular} \\ 
\hline
P14 & AI Engineer & Customer Support & \begin{tabular}[c]{@{}l@{}}Human, Social and Environmental Wellbeing \\Performance and Testing\\Reliability and Safety\end{tabular} & \begin{tabular}[c]{@{}l@{}}Unauthorized use\\Redundant actions\\Exploit trust mismatch\\AI agents' impact on jobs\\AI agents' impact on environment\end{tabular} \\ 
\hline
P15 & Technical Specialist & Legal \& Compliance & \begin{tabular}[c]{@{}l@{}}Transparency and Explainability\\Accountability\\Reliability and Safety\end{tabular} & \begin{tabular}[c]{@{}l@{}}Unexplainable and untraceable actions\\Accountability of AI agent actions\\Attack on AI agents’ external resources\\Function calling hallucination\\AI agent compliance\end{tabular} \\ 
\hline
P16 & Technical Consultant & Corporate Services & \begin{tabular}[c]{@{}l@{}}Privacy Protection and Security\\Accountability\\Transparency and Explainability\end{tabular} & \begin{tabular}[c]{@{}l@{}}Sharing IP/PI/confidential information with tools\\Sharing IP/PI/confidential information with user\\Accountability of AI agent actions\\Unexplainable and untraceable actions\\Lack of AI agent transparency\end{tabular} \\ 
\hline
P17 & Technical Solution Architect & Healthcare Services & \begin{tabular}[c]{@{}l@{}}Human, Social and Environmental Wellbeing\\Reliability and Safety\\Accountability\end{tabular} & \begin{tabular}[c]{@{}l@{}}AI agents' impact on human agency\\Accountability of AI agent actions\\Exploit trust mismatch\\Unauthorized use\\AI agents' impact on environment\end{tabular} \\ 
\hline
P18 & Data Scientist & Legal \& Compliance & \begin{tabular}[c]{@{}l@{}}Performance and Testing\\Transparency and Explainability\\Reliability and Safety\end{tabular} & \begin{tabular}[c]{@{}l@{}}Function calling hallucination\\Attack on AI agents’ external resources\\Reproducibility\\Unexplainable and untraceable actions\\Lack of AI agent transparency\end{tabular} \\ 
\hline
P19 & AI Engineer & Finance \& Banking & \begin{tabular}[c]{@{}l@{}}Performance and Testing\\Privacy Protection and Security\\Transparency and Explainability\end{tabular} & \begin{tabular}[c]{@{}l@{}}Sharing IP/PI/confidential information with user\\Sharing IP/PI/confidential information with tools\\Lack of AI agent transparency\\Mitigation and maintenance\\Incomplete AI agent evaluation\end{tabular} \\ 
\hline
P20 & Research Scientist & Sales \& Marketing & \begin{tabular}[c]{@{}l@{}}Reliability and Safety\\Performance and Testing\\Privacy Protection and Security\end{tabular} & \begin{tabular}[c]{@{}l@{}}Sharing IP/PI/confidential information with tools\\Reproducibility\\Function calling hallucination\\Sharing IP/PI/confidential information with user\\Unauthorized use\end{tabular} \\ 
\bottomrule
\end{tabular}
}
\arrayrulecolor{black}
\end{table*}

\begin{table*}
\centering
\caption{Participant Information (continued)}
\arrayrulecolor[rgb]{0.8,0.8,0.8}
\resizebox{0.8\linewidth}{!}{%
\begin{tabular}{|l|l|l|l|l|} 
\toprule
\textbf{PID} & \textbf{Job title} & \begin{tabular}[c]{@{}l@{}}\textbf{Application domain }\\\textbf{of the agentic AI products}\end{tabular} & \textbf{Selected AI ethic principles} & \textbf{Ranked agentic AI risks} \\ 
\hline
P21 & Technical Lead & Customer Support & \begin{tabular}[c]{@{}l@{}}Performance and Testing\\Accountability\\Privacy Protection and Security\end{tabular} & \begin{tabular}[c]{@{}l@{}}Sharing IP/PI/confidential information with tools\\Incomplete AI agent evaluation\\Accountability of AI agent actions\\Sharing IP/PI/confidential information with user\\Mitigation and maintenance\end{tabular} \\ 
\hline
P22 & AI Engineer & Customer Support & \begin{tabular}[c]{@{}l@{}}Performance and Testing\\Transparency and Explainability\\Reliability and Safety\end{tabular} & \begin{tabular}[c]{@{}l@{}}Unexplainable and untraceable actions\\Unauthorized use\\Reproducibility\\Lack of AI agent transparency\\Mitigation and maintenance\end{tabular} \\ 
\hline
P23 & Technical Specialist & Finance \& Banking & \begin{tabular}[c]{@{}l@{}}Transparency and Explainability\\Reliability and Safety\\Accountability\end{tabular} & \begin{tabular}[c]{@{}l@{}}AI agent compliance\\Unexplainable and untraceable actions\\Accountability of AI agent actions\\Exploit trust mismatch\\Function calling hallucination\end{tabular} \\ 
\hline
P24 & AI Engineer & Customer Support & \begin{tabular}[c]{@{}l@{}}Transparency and Explainability\\Privacy Protection and Security\\Fairness\end{tabular} & \begin{tabular}[c]{@{}l@{}}Lack of AI agent transparency\\Introduce data bias\\Sharing IP/PI/confidential information with user\\Sharing IP/PI/confidential information with tools\\Discriminatory actions\end{tabular} \\ 
\hline
P25 & AI Engineer & Information Technology & \begin{tabular}[c]{@{}l@{}}Performance and Testing\\Privacy Protection and Security\\Transparency and Explainability\end{tabular} & \begin{tabular}[c]{@{}l@{}}Mitigation and maintenance\\Incomplete AI agent evaluation\\Lack of AI agent transparency\\Unexplainable and untraceable actions\\Sharing IP/PI/confidential information with tools\end{tabular} \\ 
\hline
P26 & Technical Solution Architect & Government \& Public Sector & \begin{tabular}[c]{@{}l@{}}Human, Social and Environmental Wellbeing\\Reliability and Safety\\Accountability\end{tabular} & \begin{tabular}[c]{@{}l@{}}Function calling hallucination\\Attack on AI agents’ external resources\\AI agents' impact on human agency\\AI agents' impact on environment\\Accountability of AI agent actions\end{tabular} \\ 
\hline
P27 & Software Developer & Customer Support & \begin{tabular}[c]{@{}l@{}}Accountability\\Performance and Testing\\Privacy Protection and Security\end{tabular} & \begin{tabular}[c]{@{}l@{}}Sharing IP/PI/confidential information with user\\Incomplete AI agent evaluation\\Accountability of AI agent actions\\Sharing IP/PI/confidential information with tools\\Redundant actions\end{tabular} \\ 
\hline
P28 & Technical Solution Architect & Corporate Services & \begin{tabular}[c]{@{}l@{}}Accountability\\Performance and Testing\\Transparency and Explainability\end{tabular} & \begin{tabular}[c]{@{}l@{}}Lack of AI agent transparency\\Accountability of AI agent actions\\Unexplainable and untraceable actions\\Reproducibility\\Redundant actions\end{tabular} \\ 
\hline
P29 & Technical Specialist & Finance \& Banking & \begin{tabular}[c]{@{}l@{}}Transparency and Explainability\\Performance and Testing\\Privacy Protection and Security\end{tabular} & \begin{tabular}[c]{@{}l@{}}Sharing IP/PI/confidential information with user\\Unexplainable and untraceable actions\\Lack of AI agent transparency\\Incomplete AI agent evaluation\\Mitigation and maintenance\end{tabular} \\ 
\hline
P30 & Demand Strategist & Sales \& Marketing & \begin{tabular}[c]{@{}l@{}}Performance and Testing\\Reliability and Safety\\Privacy Protection and Security\end{tabular} & \begin{tabular}[c]{@{}l@{}}Redundant actions\\Function calling hallucination\\Reproducibility\\Attack on AI agents’ external resources\\Mitigation and maintenance\end{tabular} \\ 
\hline
P31 & Research Scientist & Corporate Services & \begin{tabular}[c]{@{}l@{}}Reliability and Safety\\Performance and Testing\\Accountability\end{tabular} & \begin{tabular}[c]{@{}l@{}}AI agent compliance\\Incomplete AI agent evaluation\\Accountability of AI agent actions\\Mitigation and maintenance\\Redundant actions\end{tabular} \\ 
\hline
P32 & Research Scientist & Sales \& Marketing & \begin{tabular}[c]{@{}l@{}}Accountability\\Performance and Testing\\Privacy Protection and Security\end{tabular} & \begin{tabular}[c]{@{}l@{}}Sharing IP/PI/confidential information with user\\Sharing IP/PI/confidential information with tools\\AI agent compliance\\Redundant actions\\Reproducibility\end{tabular} \\ 
\hline
P33 & AI Engineer & Government \& Public Sector & \begin{tabular}[c]{@{}l@{}}Transparency and Explainability\\Performance and Testing\\Human-centred Values\end{tabular} & \begin{tabular}[c]{@{}l@{}}Redundant actions\\Reproducibility\\Incomplete AI agent evaluation\\Unexplainable and untraceable actions\\Over- or under-reliance on AI agents\end{tabular} \\ 
\hline
P34 & Research Scientist & Information Technology & \begin{tabular}[c]{@{}l@{}}Privacy Protection and Security\\Performance and Testing\\Accountability\end{tabular} & \begin{tabular}[c]{@{}l@{}}Sharing IP/PI/confidential information with tools\\Sharing IP/PI/confidential information with user\\Accountability of AI agent actions\\AI agent compliance\\Mitigation and maintenance\end{tabular} \\ 
\hline
P35 & Research Scientist & Information Technology & \begin{tabular}[c]{@{}l@{}}Privacy Protection and Security\\Reliability and Safety\\Accountability\end{tabular} & \begin{tabular}[c]{@{}l@{}}Sharing IP/PI/confidential information with user\\AI agent compliance\\Sharing IP/PI/confidential information with tools\\Exploit trust mismatch\\Attack on AI agents’ external resources\end{tabular} \\
\bottomrule
\end{tabular}
}
\arrayrulecolor{black}
\end{table*}

\end{document}